\documentclass{IEEEtran}
\usepackage{cite}
\usepackage{amsmath,amssymb,amsfonts}
\usepackage{algorithmic}
\usepackage{graphicx}
\usepackage{textcomp}
\usepackage{multirow}
\usepackage{color}

\def\BibTeX{{\rm B\kern-.05em{\sc i\kern-.025em b}\kern-.08em
    T\kern-.1667em\lower.7ex\hbox{E}\kern-.125emX}}
\begin{document}
\title{Measurements and Modeling of Air-Ground Integrated Channel in Forest Environment Based on OFDM Signals}
\author{Zhe Xiao,        
        Shu Sun,~\IEEEmembership{Member,~IEEE,}        
        Na Liu,~\IEEEmembership{Member,~IEEE,}
        
        Lianming Xu,
        Li Wang,~\IEEEmembership{Senior Member,~IEEE}
\thanks{This work was supported in part by the Fundamental Research Funds for the Central Universities under Grant No. 24820232023YQTD01, in part by the National Natural Science Foundation of China under Grants U2066201, 62101054, and 62271310, in part by the Double First-Class Interdisciplinary Team Project Funds 2023SYLTD06, and in part by Key Laboratory of Satellite Navigation Technology under Grant WXDHS2023105.}}

\maketitle

\begin{abstract}
Forests are frequently impacted by climate conditions, vegetation density, and intricate terrain and geology, which contribute to natural disasters. Personnel engaged in or supporting rescue operations in such environments rely on robust communication systems to ensure their safety, highlighting the criticality of channel measurements in forest environments. However, according to current research, there is limited research on channel detection and modeling in forest areas in the existing literature. This paper describes the channel measurements campaign of air and ground in the  Arxan National Forest Park of Inner Mongolia. It presents measurement results and propagation models for ground-to-ground (G2G) and air-to-ground (A2G) scenarios. The measurement campaign uses orthogonal frequency division multiplexing signals centered at 1.4 GHz for channel sounding. In the G2G measurement, in addition to using omnidirectional antennas to record data, we also use directional antennas to record the arrival angle information of the signal at the receiver. In the A2G measurement, we pre-plan the flight trajectory of the unmanned aerial vehicle so that it can fly at a fixed angle relative to the ground. We present path loss models suitable for G2G and A2G in forest environments based on the analysis of measurement results. The results indicate that the proposed model reduces error margins compared with other path loss models. Furthermore, we derive the multipath model expression specific to forest environments and conduct statistical analysis on key channel parameters e.g., shadow fading factor, root mean square delay spread, and Rician K factor. Our findings reveal that signal propagation obstruction due to tree crowns in A2G communication is more pronounced than tree trunk obstructions in G2G communication. Adjusting the elevation angle between air and ground can enhance communication quality. These results and models serve as valuable references for future research on forest channel characteristics.
\end{abstract}


\section{Introduction}
\label{sec:introduction}
\IEEEPARstart{F}orest environments are characterized by variable climates, dense vegetation, and complex terrain, making them susceptible to natural disasters such as fires, floods, and earthquakes. In recent years, both natural disaster prevention and rescue have put forward increasingly high requirements for relevant departments to handle emergency events. Therefore, building an efficient and applicable emergency communication system to provide more timely rescue services for forest areas has become an important and urgent challenge. 

In modern communication systems, studying the characteristics of different frequency bands is crucial. Low-frequency bands (e.g., HF and VHF) offer excellent penetration and longer range but have limited bandwidth and lower data transfer rates \cite{vhf}. Mid-frequency bands (e.g., UHF) provide a broader bandwidth suitable for high-speed data transmission but are prone to multipath effects and interference \cite{uhf}. High-frequency bands (e.g., millimeter waves) enable extremely high data transfer rates, ideal for high-speed communication \cite{sun_tap,sun_jsac}, but have shorter range and are more sensitive to weather conditions and obstacles.
Research on wireless channels covers many fields, including mobile communication \cite{2,6,7}, satellite communication \cite{3,39}, radar \cite{4}, and wireless sensor networks \cite{5}. Most existing channel measurement activities revolve around cities\cite{8}, suburbs \cite{suburb}, rural areas \cite{9}, oceans \cite{10,40}, etc. Unlike other environments, the channels in forest areas are more complex, and different natural landforms, vegetation types, vegetation distribution characteristics, temperature, and humidity \cite{12,13,14,15,16,17} can all affect signal transmission. The dense vegetation, complex terrain, and dangerous organisms in forests pose significant challenges to channel-sounding activities, resulting in limited channel measurements in forest environments.

Due to the harsh environment in forest areas, many scholars use unmanned aerial vehicles (UAVs) as aerial base stations (BSs) to conduct channel measurements. UAVs are promising temporary service BSs that can provide communication coverage and increase network capacity \cite{uav_serve}. Therefore, studying the channel between UAVs and ground nodes in forest areas has become a hot topic. There are numerous previous studies on air-to-ground (A2G) channel modeling. In \cite{28}, the authors described simultaneous dual frequency (L-band~970 MHz, C-band~5 GHz) measurement activities and over-water measurement scenarios and provided the results of path loss and root mean square delay spread (RMS-DS), as well as the results of channel stationary distance (SD), for calculating the correlation between the Rician K-factor and the two receiving antennas we used in each frequency band. In \cite{29}, the authors analyzed UAV's channel narrowband and wideband characteristics to ground links under medium elevation, and line-of-sight (LoS) and non-line-of-sight (NLoS) conditions. Using ray tracing (RT), the authors investigated the received power, time-varying channel impulse response (CIR), average time of arrival (ToA), and delay characteristics along the UAV trajectory. In \cite{30}, the authors propose channel measurement activities and conduct a comprehensive study on the parameters of A2G channels in different environments in Pakistan. The measured channel parameters are provided based on the empirical results of control path loss, RMS-DS, and multipath components (MPCs). However, due to signal propagation through the canopy layer between air and ground, and varying attenuation caused by different elevation angles between air and ground, the communication quality of aerial-based stations may not necessarily be superior to that of ground-based stations. Currently, there is a lack of integrated channel measurement activities between air and ground.

At present, wireless channel models can be mainly categorized as statistical channel models, semi-deterministic channel models, and deterministic channel models. The semi-deterministic channel model is represented by the Geometry-Based Stochastic Model (GBSM), which assumes that the distribution of scatterers has a certain regularity. However, the GBSM method remains incapable of calculating precise channel transmission characteristics for specific scenarios \cite{18}. The deterministic channel model can calculate the field strength of any point within a particular scene, mainly by using RT methods. In \cite{19}, the authors propose extending a 2-D RT-based model for radio wave propagation to encompass real-sized trees and outdoor forest scenes. However, to ensure the accuracy of the results, RT algorithms require precise 3D models of the environment and a significant amount of simulation time. Statistical channel modeling combines practical tests with probability and statistical theory to summarize and generalize channel characteristics. It simplifies computational complexity, offers broad applicability, and provides a comprehensive definition of channel parameters. Standard propagation models such as the Okumura-Hata model \cite{20}, the Cost-231 Hata model \cite{21}, and the standard propagation model \cite{22} are widely used. The Okumura-Hata model, for instance, leverages diverse frequency measurements across various terrains to depict field strength-distance relationships. However, its complexity poses challenges in practical use. The Erceg model \cite{23}, proposed in 2010, classifies terrain into three categories and provides tailored channel characteristics. Similarly, the Stanford University Interim (SUI) channel model \cite{24}, introduced in 2010, offers six-channel models representing various terrain types and conditions typical in the continental United States. Though initially developed for North American environments, these models provide valuable insights into channel behavior.

However, forests are a unique environment where only a few inches of movement may cause signal propagation condition changes from LoS to NLoS and vice versa. The transformation range of the Rayleigh factor is relatively large, ranging from -30 dB to +10 dB \cite{25}\cite{26}. The small-scale fading of forests also depends on factors such as frequency, antenna height, forest density, etc. The combination of the tree trunk, crown, leaves, and shrubs makes the multipath effect of forest environments more pronounced. Multipath components can be generated by the specular reflection of tree trunks, as well as by the scattering from tree crowns or shrubs. The time different paths of signals take from different paths to reach the receiver (RX) is different, which increases the delay spread. Therefore, the technical difficulties of forest channels mainly include signal propagation changes caused by movement, widespread variations in the Rayleigh factor, the complexity of small-scale fading, and the pronounced presence of multipath effects. These factors render signal transmission and reception in forest environments more complex and challenging.

To tackle the challenges above, this paper conducts channel measurement campaigns based on OFDM signals, analyzes the large-scale and small-scale fading of channels in forest environments based on the measurement results, and proposes a channel model suitable for forest areas. The main contributions of this paper are summarized as follows:

\begin{itemize}
    \item We report the channel measurement results of ground-to-ground (G2G) and A2G in two forest environments at the 1.4 GHz carrier frequency.
    \item Based on measured data, we establish path loss models for G2G and A2G to characterize the large-scale fading. Results show that the proposed path loss model has small errors and better performance than existing classical models for forest environments.
    \item Based on measured data, we perform statistical analysis and propose a multipath propagation model to characterize the small-scale fading. We compare and analyze the RMS-DS, angle spread (AS), and Rician K factor of G2G and A2G channels.
\end{itemize}

The remainder of this paper is organized as follows. In Section II, we introduce the structure and formula of OFDM signals used for measurement. Section III introduces the methods and steps of channel measurement. Section IV proposes path loss models for G2G and A2G channels in forest areas. Section V introduces the extraction and statistics of small-scale parameters. Section VI proposes channel models for G2G and A2G in forest areas. Finally, Section VII concludes this paper.

\section{The Application Of OFDM Signal In Channel Measurement}
This section introduces the characteristics of the OFDM signal, discusses the clock synchronization of the measurement equipment, and derives the corresponding expressions for the CIR.

\subsection{Channel Sounding Signal}
In recent years, OFDM has been widely adopted in wireless communication systems due to its high data rate, efficient bandwidth utilization, and robustness to multipath delays. This channel measurement uses LTE (Long Term Evolution) OFDM signals \cite{31} with a bandwidth of 20 MHz, and an effective bandwidth of 18 MHz, where the remaining 2 MHz serves as the protection bandwidth. Fig. 1 shows the frame structure of the OFDM signal. The subcarrier spacing is 15 kHz, and the number of subcarriers is 1200. Each frame has 14 symbols and consists of a cyclic prefix and data, where the length of the data depends on the number of Inverse Fast Fourier Transform (IFFT) points. The cyclic prefix length of symbols 0 and 7 is 160, while the cyclic prefix length of the other symbols is 144. Because the selected number of IFFT points is 2048, although the number of input subcarriers is 1200 (the remaining 848 are not used), the output is 2048 points. Therefore, the interval between sampling points is 32.55 ns, and the sampling frequency is 30.72 MHz. Symbol 3 and Symbol 10 are pilot symbols in the frame structure.

\begin{figure}[htbp]
\centerline{\includegraphics[width=8.5cm]{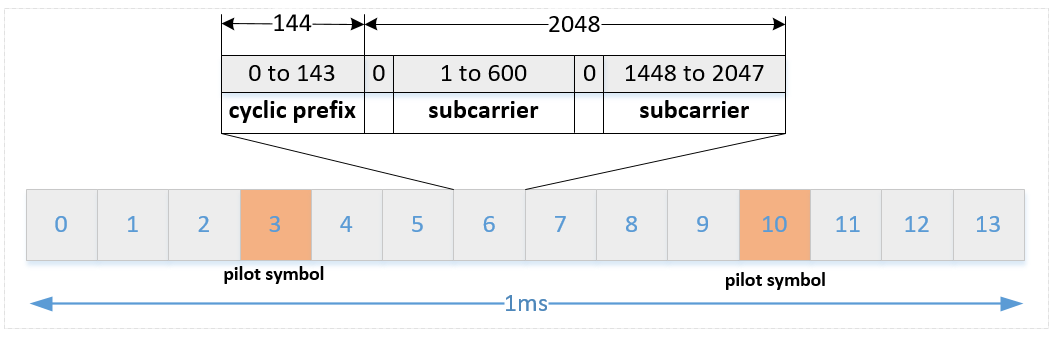}}
\caption{OFDM signal frame structure.}
\label{fig}
\end{figure}

In channel measurement, precise clock synchronization between the TX and RX is crucial for accurate delay measurement. Our equipment integrates a GPS module and uses the Zadoff-Chu (ZC) sequence for the time-frequency synchronization of OFDM signals. Fig. 2 illustrates the synchronization process between the RX and TX. The GPS RX provides a 1 Pulse Per Second (1PPS) signal for device synchronization, relying on atomic clocks known for their precision. The 1PPS signal, with a pulse width of 200-300 ms, is a highly accurate synchronization signal, initiating Time of Day (TOD) data transmission. The OFDM frame structure, with 14 symbols, ensures 30720 valid data points. We extend the signal length to 40000 points to compensate for internal chip delays. 

\begin{figure}[htbp]
\centerline{\includegraphics[width=8.5cm]{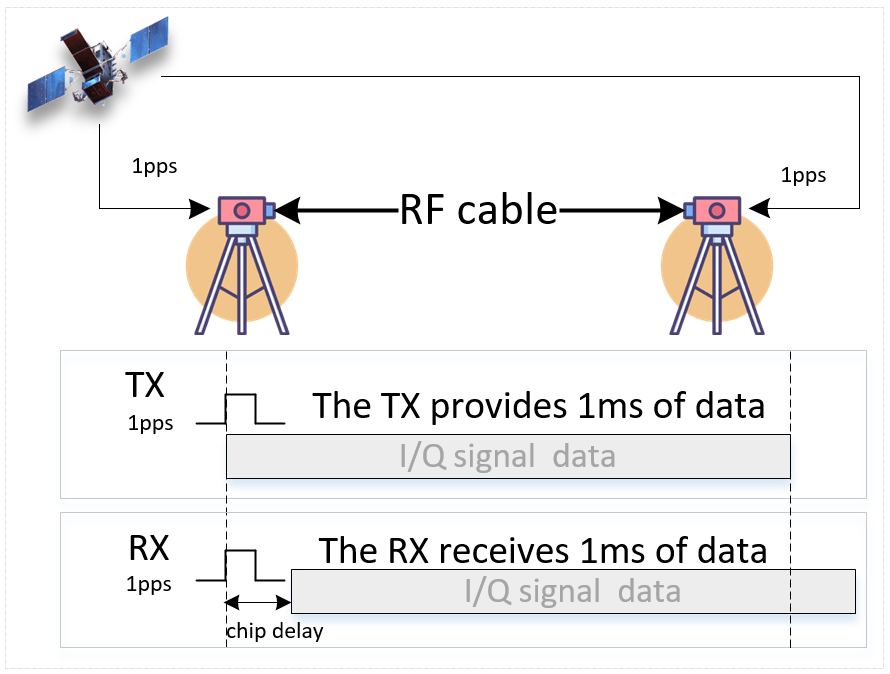}}
\caption{OFDM signal synchronization process.}
\label{fig}
\end{figure}

\subsection{Analysis of Channel Impulse Response}
Due to the presence of time-dispersive channels in OFDM systems, multipath delay spread generates inter-symbol interference (ISI). Additionally, subcarrier components will no longer be orthogonal, resulting in inter-subcarrier interference (ICI). Therefore, cyclic prefixes are added in practical systems to mitigate both types of interference. For OFDM transmission signals, consider an OFDM system with $N$ orthogonal subcarriers, which can be represented as:
\begin{equation}\phi_k\left(t\right)=e^{j2\pi\left(f_c+\frac{k}{T}\right)t},\end{equation}

\noindent where $f_c+\frac{k}{T}$ is the frequency of the $k$-th subcarrier, $k=0,1,2,\cdots,N-1$, $f_{c}$ represents the center carrier frequency of the OFDM signal. The symbol length after adding a cyclic prefix is $T_{W}=T+T_{G}$, where $T_{W}$ is the total length of OFDM symbols, $T_{G}$ is the length of the protection interval, $T=NT_{S}$ is the pure data length of OFDM symbols, which is also the period of Fast Fourier Transform (FFT)/IFFT. $T_{S}$ is the system sampling interval, with no oversampling $T_s{=}1/B$ , and $B$ is the system bandwidth. The  time-domain OFDM symbol (without cyclic prefix) after modulating $N$ parallel data can be expressed as
\begin{equation}s\left(t\right)=\frac{1}{\sqrt{T}}\sum_{k=0}^{N-1}d_{k}\phi_k\left(t\right),(-T_{G}\leq t\leq T),\end{equation}

\noindent where $d_{k}$ is the complex signal modulated on the $k$-th subcarrier. 

After the additive white Gaussian noise $n(t)$, the received signal $r(t)$ can be expressed as
\begin{equation}r\begin{pmatrix}t\end{pmatrix}=s\begin{pmatrix}t\end{pmatrix}*h\begin{pmatrix}t,\tau\end{pmatrix}+n\begin{pmatrix}t\end{pmatrix},\end{equation}

\noindent where $h(t,\tau)$ is the time-domain CIR of the wireless channel. Under the multipath condition, if the wireless channel remains constant for an OFDM symbol duration, the CIR can be modeled as the Saleh Valenzuela channel model \cite{salehvalenzuela}, which is written as 

\begin{equation}
\begin{aligned}
\begin{split}
&h(t)=\sum_{l=0}^L\sum_{m=0}^Ma_{l,m}(t)e^{j\phi_{l,m(t)}}\delta(t-\Gamma_l-\tau_{l,m}(t)) 
\end{split}
\end{aligned},\end{equation}

\noindent where $L$ is the total number of clusters, $M$ is the number of multipath within each cluster, $a_{l,m}$, $\mathbf{\phi}_{l,m}$, $\mathbf{\tau}_{l,m}$ are the complex fading coefficient, phase, and the relative delay of the $m$-th multipath in the $n$-th cluster, respectively, $\Gamma_{l}$ is the delay of the $n$-th cluster. In general, due to the slow channel change speed compared to the measurement time interval, the propagation delay and complex fading coefficient of multipath components in the channel can be considered constant during measurement and $\mathbf{\phi}_{l,m}$ follows a uniform distribution in the interval (0, 2$\pi$). Meanwhile, the multipath complex fading coefficients $a_{l,m}$ are independent. In single-input single-output (SISO) systems, the analysis of small-scale fading characteristics typically focuses on multipath amplitude and delay propagation, often overlooking the information in the channel angular domain. Therefore, we can rewrite (4) as follows

\begin{equation}
\begin{aligned}
\begin{split}
&h(t)=\sum_{l=0}^{L}\sum_{m=0}^{M}a_{l,m}\delta(t-\Gamma_{l}-\tau_{l,m})
\end{split}
\end{aligned},\end{equation}

the received OFDM signal $r(t)$ is written as 

\begin{equation}\begin{aligned}
&r(t)=\sum_{l=0}^L\sum_{m=0}^Ma_{l,m}s(t-\Gamma_l-\tau_{l,m})+n(t) \\
&=\frac{1}{\sqrt{T}}\sum_{k=0}^{N-1}d_k\sum_{l=0}^L\sum_{m=0}^Ma_{l,m}e^{-j2\pi(f_c+\frac{k}{T})(\tau_{l,m}+\Gamma_l)}e^{j2\pi(f_c+\frac{k}{T})t} \\
&+n(t).
\end{aligned}\end{equation}

Sampling the signal with the time interval of the symbol duration $T_s$ will generate the discrete-time samples
\begin{equation}R(k)=d_k\sum_{l=0}^L\sum_{m=0}^Ma_{l,m}e^{-j2\pi(f_c+\frac{k}{T})(\tau_{l,m}+\Gamma_l)}+N(k).\end{equation}

Then, it is easy to know that the channel frequency domain response of the $k$-th subcarrier is
\begin{equation}x(k)=\sum_{l=0}^L\sum_{m=0}^Ma_{l,m}e^{-j2\pi(f_c+\frac{k}{T})(\tau_{l,m}+\Gamma_l)})+N(k).\end{equation}

Finally, we can obtain the CIR of the $k$-th subcarrier by performing IFFT on CFR.

\section{Measurement Setup}
In this section, we introduce measurement procedures. The measurement scenario is divided into G2G and A2G, as shown in Fig. 3, conducted in two forest areas with different vegetation distributions. In addition, we use directional antennas to measure the angular diffusion in forest areas.

\subsection{Description of Measurement Locations}

The channel measurement took place in Arxan National Forest Park. The park lies in the southwest foothill of the volcanic mountain range of the Greater Khingan. The park is located in the continental climate zone of the Mongolian Plateau, belonging to the cold temperate humid zone. The annual average temperature is -3.2 ℃, with an average yearly temperature difference of 42 ℃ and an average precipitation of 45.1 mm. The vegetation type belongs to the mixed coniferous and broad-leaved forest in the cold temperate zone, with a forest coverage rate of 80\%. The tree can reach a height of up to 25 m and a diameter at breast height of up to 50 cm. The Arxan National Forest Park has various original ecological attractions such as primitive forests, volcanic ruins, hot springs, rivers and lakes, and the Grand Canyon. Therefore, the terrain also undergoes significant changes.
The measurement was conducted in two forest areas near the Dichi, primarily composed of deciduous larch and birch forests. The measurement point is situated in a flat area with minimal terrain change (±0.5 m). The measurement points are categorized into two different forest vegetation compositions, as shown in Fig. 4. UAVs were used to take aerial photographs of the measurement locations from various heights and angles to observe the forest structure. The distance between trees in Fig. 4 (a) and (b), mainly larch, is about 2-5 m, with trunk diameters up to 50 cm and heights up to 30 m. There are many weeds between the trees. In Fig. 4 (c) and (d), which depict white birches, the distance between trees is about 1-3 m, with trunk diameters at breast height reaching 25 cm and tree heights generally around 25 m. The space between these trees contains many weeds and, unlike the deciduous larch forest, also features many shrubs. Table \uppercase\expandafter{\romannumeral1} provides channel measurement parameters, including the longitude and latitude of the measured locations, transmission and reception equipment, antenna type, and more.

\begin{figure}[htbp]
\centerline{\includegraphics[width=0.5\textwidth]{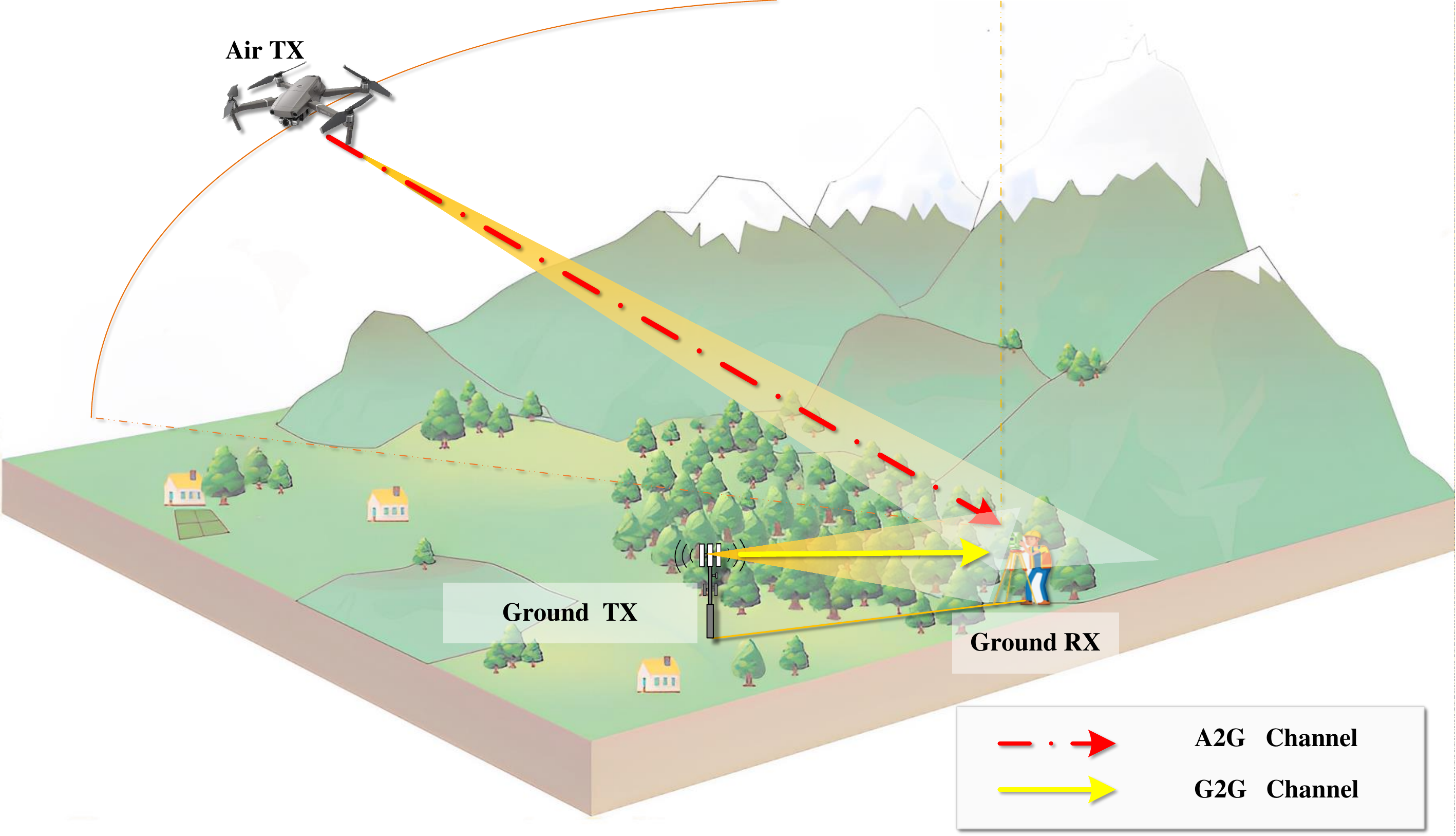}}
\caption{Channel measurement scenario.}
\label{fig}
\end{figure}

\begin{figure*}[htbp]
 \centering
	\begin{minipage}{0.45\linewidth}
		\centerline{\includegraphics[width=\textwidth]{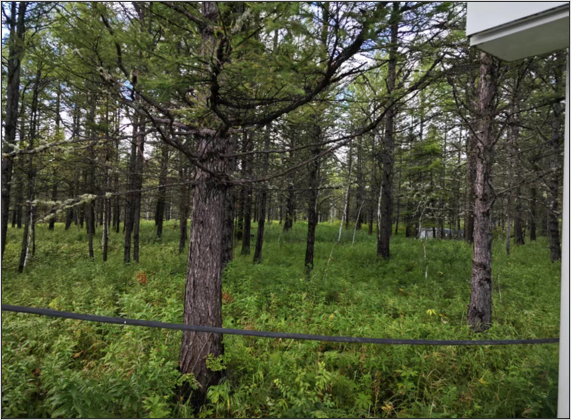}}
	 \centerline{(a)}
	\end{minipage}
	\begin{minipage}{0.45\linewidth}
		\centerline{\includegraphics[width=\textwidth]{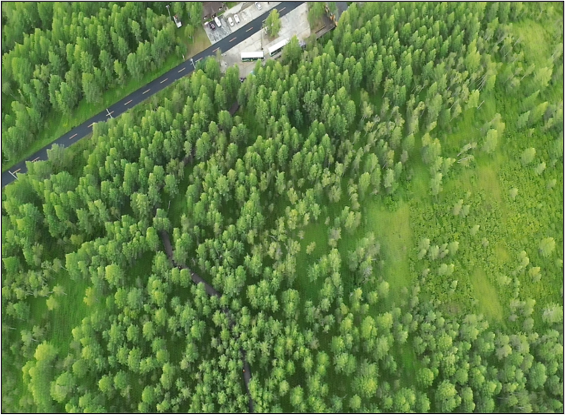}}
        \centerline{(b)}
	\end{minipage} \\
    \begin{minipage}{0.45\linewidth}
		\centerline{\includegraphics[width=\textwidth]{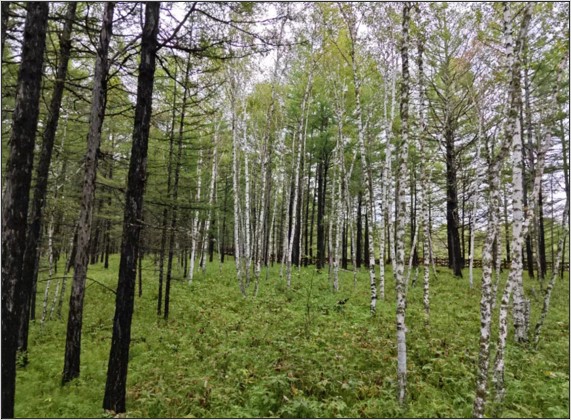}}
        \centerline{(c)}
	\end{minipage} 
   \begin{minipage}{0.45\linewidth}
		\centerline{\includegraphics[width=\textwidth]{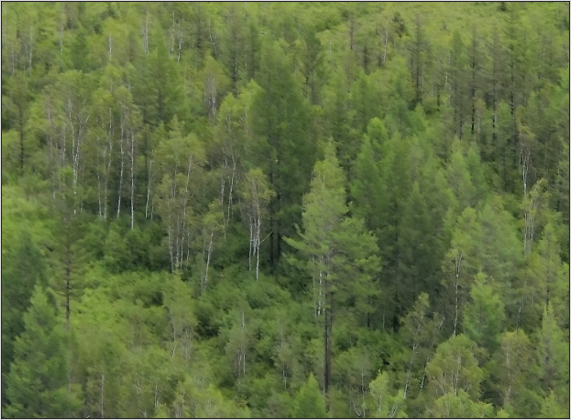}}
        \centerline{(d)}
	\end{minipage} 
        
    \caption{Two measurement scenarios: (a) Larch forest. (b) Aerial photographs of UAV with long-range top view in larch forest.
(c) White Birch forest. (d) Aerial photographs of UAV with short-range Oblique view in white birch Forest}
	\label{}
\end{figure*}

\subsection{Description of G2G Measurement Procedure}
We installed omnidirectional fiberglass antennas at the RX and TX to obtain path loss and multipath information. Before the measurement, we planned several travel routes and measurement points for the RX. The TX and RX were placed in the forest area, and calibrations were performed in the far field of the transmit antenna to ensure that the system functioned correctly and to obtain the system gain. Then, the initial received signal strength (RSS) was recorded under the LoS condition of the RX and TX. Afterward, the TX remains stationary and sends continuous signals to the RX. The RX moves towards the forest's depths in a specific direction. At this time, the RX received and recorded the RSS at intervals of $t$ and, simultaneously, recorded the GPS information at this position at intervals of $t$. After the RX reaches the designated location, it collects and records 40000 points of the signal. As the distance between the RX and the TX increases until communication between the RX and TX is interrupted, data collection is completed, and the GPS position information, RSSI information, and OFDM data obtained by the RX are saved. Fig. 5 (a) shows the equipment installing omnidirectional antennas. Fig. 6 shows the route and region of measurement in the forest area.

We installed directional antennas at the RX to obtain parameters such as angular power spectrum (APS) and azimuth angular spread (AS). During the measurement, we fix the TX, and the RX remains fixed after reaching the designated position. After rotating the half-power beamwidth of the flat antenna at the RX, the personnel record the RSSI and 40000 points of the received signal. The half-power beamwidth of the antenna we use is 30°, so after 12 rotations, we can measure RSSI and signal data in different directions for one cycle. Fig. 5 (b) shows the equipment installing directional antennas.

\begin{table}[htbp]
\centering
\caption{MEASUREMENT PARAMETERS}
\resizebox{1.0\linewidth}{!}{
\begin{tabular}{ll}
\hline
\textbf{Parameter}                  & \textbf{Value}                   \\ \hline
TX$\backslash$RX      & LENA P35                \\
Coordinate of Larch forest & (47.307141, 120.476046) \\
Coordinate of Birch forest & (47.308549, 120.477147) \\
Measurement type           & SISO                    \\
Ground TX Antenna Height   & 1.8 m                    \\
Ground RX Antenna Height   & 1.8 m                    \\
TX Power                   & 27 dBm                   \\
Omnidirectional Gain       & 8 dBi$\backslash$3 dBi                    \\
Directional Antenna Gain       & 16 dBi                    \\
Center Frequency           & 1.4 GHz                    \\
Bandwidth                  & 20 MHz                   \\ \hline
\end{tabular}}
\end{table}

\begin{figure}[htbp]
 \centering
	\begin{minipage}{0.48\linewidth}
		\centerline{\includegraphics[width=\textwidth]{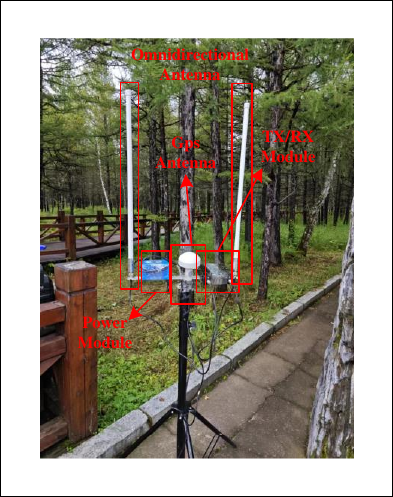}}
	 \centerline{(a)}
	\end{minipage} 
	\begin{minipage}{0.48\linewidth}
  \centering
		\centerline{\includegraphics[width=\textwidth]{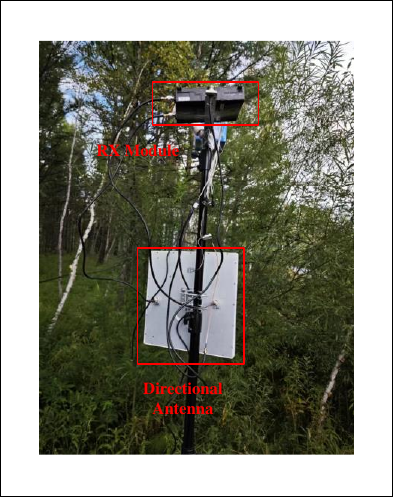}}
        \centerline{(b)}
	\end{minipage} 
        
\caption{Channel measurement equipment: (a) The ground signal transmission device with omnidirectional antennas. (b) The ground signal receiving device with directional antennas.}
	\label{}
\end{figure}

\begin{figure}[htbp]
\centerline{\includegraphics[width=0.5\textwidth]{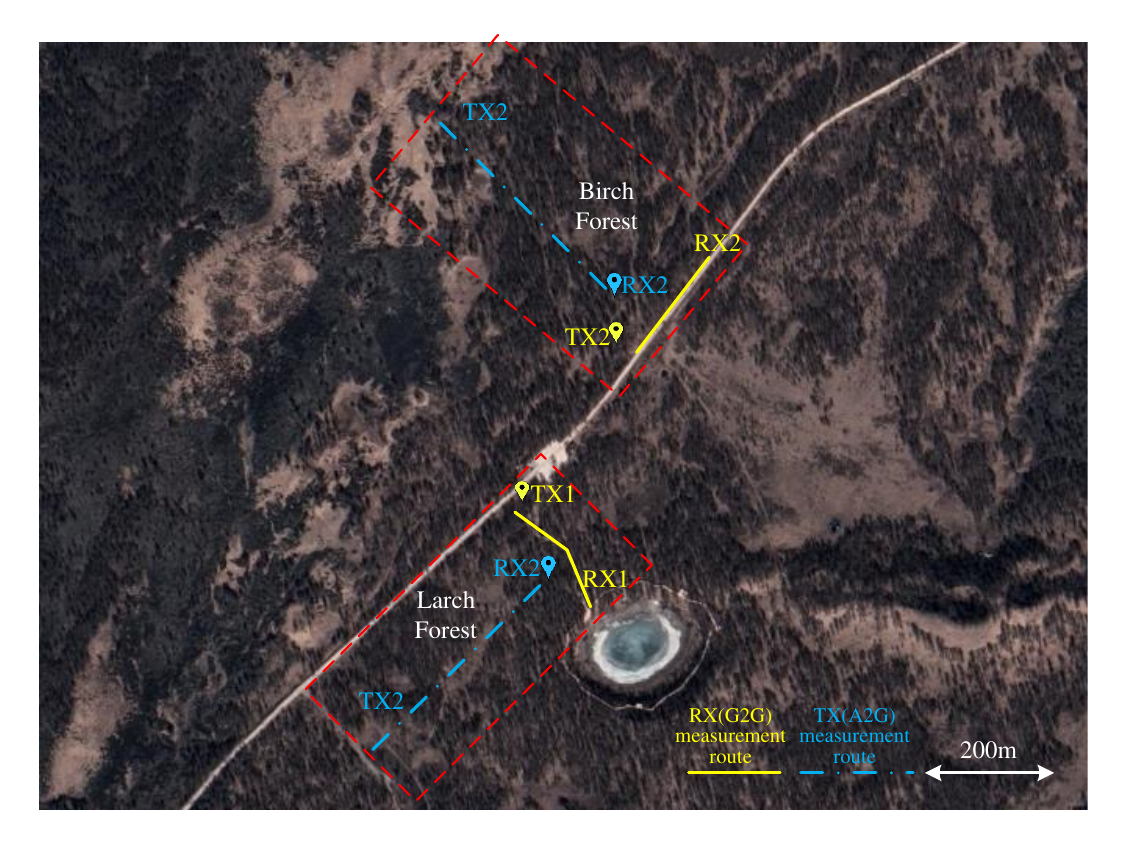}}
\caption{Channel measurement environment.}
\label{fig}
\end{figure}

\subsection{Description of A2G Measurement Procedure}

The study of A2G channels in forest areas generally requires the use of UAVs. Therefore, we attach measurement equipment to the UAV, as the aerial TX, while the RX is deployed on the ground and equipped with omnidirectional antennas to receive signals, as shown in Fig. 7. Afterward, we traced the UAV's trajectory over the larch and birch forests. Three tracks were drawn for each forest area to study the relationship between the A2G channels and the elevation angles corresponding to 30 °, 60 ° and 90 °. Fig. 6 shows the trajectory of the UAV. The steps of the A2G channel measurement are similar to that of the G2G channel measurement, where the TX continuously sends continuous signals, and the RX records the received RSSI and current time every $t$ seconds. When the UAV flies to the designated point, the RX collects the signal. Calculate the distance between the receiving and transmitting terminals based on the GPS position and time uploaded by the UAV.

\begin{figure}[htbp]
 \centering
\centerline{\includegraphics[width=0.48\textwidth]{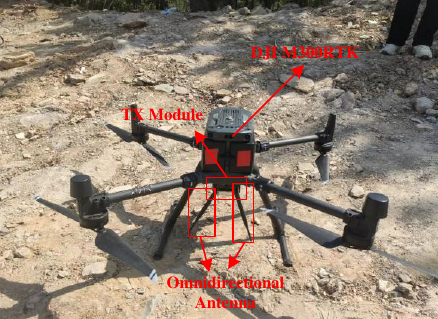}}
\caption{The aerial signal transmission device.}
\label{fig}
\end{figure}

\section{Large-Scale Fading}

In this section, we will study the large-scale propagation of signals in forest areas, including A2G and G2G, propose a path loss model for forest channels, and compare our results with several standard models proposed in the literature.

\subsection{G2G path loss model}
We adopt some empirical path loss models as the baseline generic model, which are the close-in free-space reference distance (CI) model \cite{32}\cite{33} and the free-space path loss (FSPL) model. The expressions of the two models are as follows:

\begin{small}
\begin{equation}
\mathrm{P L}_{\mathrm{CI} }=10 n \log_{10}{\frac{d}{d_{0} } }+20\log_{10}{\left ( \frac{4\pi\times 10^{9}}{c}\right )}+20 \log_{10}{f} ,
\end{equation}
\end{small}

\begin{small}
\begin{equation}
\mathrm{P L}_{\mathrm{FSPL} }=20\log_{10}{(\frac{4\pi fd\times 10^{9}}{c}  ) } ,
\end{equation}
\end{small}

\noindent where $n$ denotes the path loss exponent (PLE), $d_{0}$ is the close-in free-space reference distance and is set to 1 m \cite{32}, $d$ is the 3-D T-R separation distance in meters, $f$ is the carrier frequency in GHz,  $c$ is the speed of light in meters.

In addition, the ITU horizontal forest model \cite{34}, as an additional path loss model, can reflect the attenuation of signals caused by vegetation in forest areas. The expression for the ITU horizontal forest model is as follows:

\begin{small}
\begin{equation}
\mathrm{P L}_{\mathrm{ITU-H} }=A_m\left[ 1-e^ {\left( -d\mu/A_m \right)} \right] ,
\end{equation}
\end{small}

\noindent where $\mu$ denotes the specific attenuation for very short vegetative paths (dB/m) and $A_m$ denotes the maximum attenuation for one terminal within a specific type and depth of vegetation (dB). Therefore, we combine the FSPL model with the ITU horizontal forest model as the total path loss \cite{35} caused by the forest area. The expression for the entire path loss model FSPL-H is as follows:

\begin{small}
\begin{equation}
\mathrm{P L}_{\mathrm{FSPL-H} }=20\log_{10}{ (\frac{4\pi fd\times 10^{9}}{c}  )} +A_m\left[ 1-e^ {\left( -d\mu/A_m \right)} \right] .
\end{equation}
\end{small}

We also used the SUI model introduced in the introduction as a comparison, which is ideal for forest environments in North America. Its expression is as follows:
\begin{small}
\begin{equation}\mathrm{P L}_{\mathrm{SUI} }=A+10\gamma\log_{10}\biggl(\frac{d}{d_0}\biggr)\quad\text{for }d>d_0\end{equation}
\end{small}


\noindent where $\mathrm{A}=20\log_{10}(4\pi\mathrm{d}_{0}/\lambda)$, $\lambda$ is the wavelength in m. $\gamma$ is the path-loss exponent with $\gamma=(\mathrm{a}-\mathrm{b}\mathrm{h_b}+\mathrm{c}/\mathrm{h_b})$, $h_b$ is the height of the base station in m, $d_0 = 100m$ and a,b,c are constants dependent on the terrain category given in Table \uppercase\expandafter{\romannumeral2} and reproduced below. For distances less than 100 m, the path loss can be considered free space loss.

In our recent work, we conducted field data measurements in the forest areas of Jiaozi Snow Mountain and Pudu-river Dry-hot Valley in Yunnan Province. We proposed a new path loss model, the Beijing University of Posts and Telecommunications horizontal forest (BHF) model \cite{36}. As signal propagation distance increases, the attenuation in dense vegetation becomes non-linear and approaches an upper limit. When the input approaches positive infinity, the output of the $tanh$ function asymptotically converges to 1, making it suitable for simulating attenuation caused by vegetation. The expression of this model is as follows:

\begin{small}
\begin{equation}
\mathrm{P L}_{\mathrm{BHF}}=10\alpha  \log_{10}{d}+\beta+\zeta \tanh (d /20)+20 \log_{10}{f} ,
\end{equation}
\end{small}

\noindent where $\alpha$ is a coefficient showing the dependence of path loss on the conventional log-scaled distance, $\beta$ is an optimized offset value for path loss in decibels, $\zeta$ is a coefficient characterizing the path loss caused by vegetation attenuation. The attenuation caused by vegetation gradually increases with increasing distance and approaches an upper bound. Therefore, we use the $tanh$ (hyperbolic tangent) function to express the additional attenuation caused by vegetation on the signal.

Through data analysis, we found that at short distances, vegetation-induced attenuation is minimal, and propagation is LoS, dominated by free-space loss. As distance increases, the measurement scene gradually becomes NLOS, where vegetation obstruction becomes significant, resulting in higher path loss. Therefore, we designed a piecewise function to capture this process, which is commonly employed in studies \cite{fenduan1}\cite{fenduan2}. We made modifications to the BHF model by using a parameter $d_0$ equal to 30 m as the interval point for the distance, which is expressed as follows:

\begin{small}
\begin{equation}\mathrm{P L}_{\mathrm{BHF-M}}=\begin{cases}10n\log_{10}(d/10)+20\log_{10}f+m  ,\text{ }d\leq d_0\\10\alpha\log_{10}((d-d_0)/10+1)+\beta+\\ \qquad
\zeta\tanh((d-d_0)/20)+ PL_{d=d_0} ,\text{ }d>30,&\end{cases}\end{equation}
\end{small}

\noindent where $d$ is the distance between the TX and RX in meters, $n$ is the coefficient that represents the correlation between path loss and conventional logarithmic scale distance, and $m$ is the optimized offset value of path loss in dB; $\alpha$ is the coefficient that represents the correlation between path loss and conventional logarithmic scale distance, $\beta$ is the optimized offset value of path loss in dB, and $\zeta$ is the coefficient that characterizes the path loss caused by vegetation attenuation.

\begin{figure}[h]
	\begin{minipage}{0.96\linewidth}
		\centerline{\includegraphics[width=\textwidth]{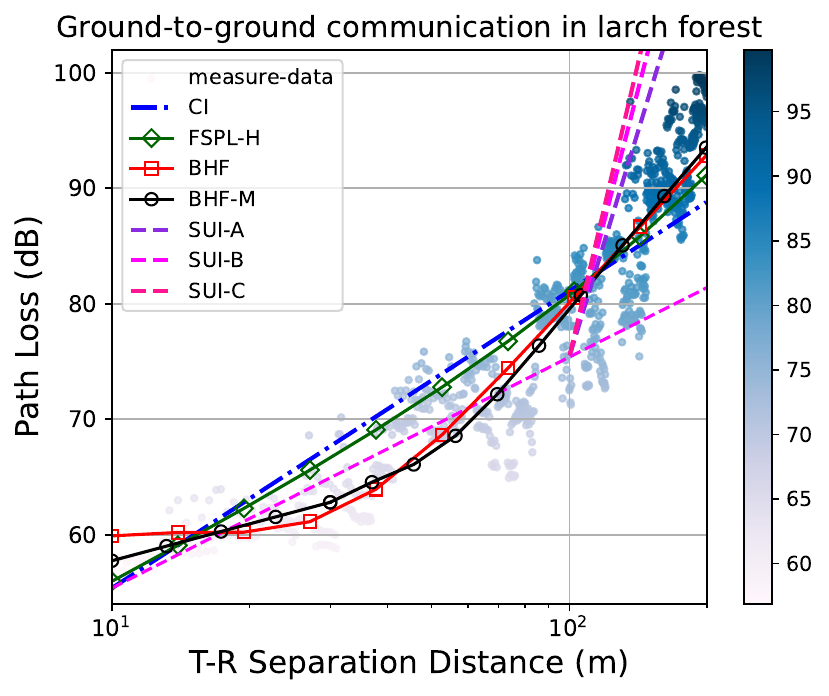}}
        \centerline{(a)}
	\end{minipage} \\
    \begin{minipage}{0.96\linewidth}
        \centerline{\includegraphics[width=\textwidth]{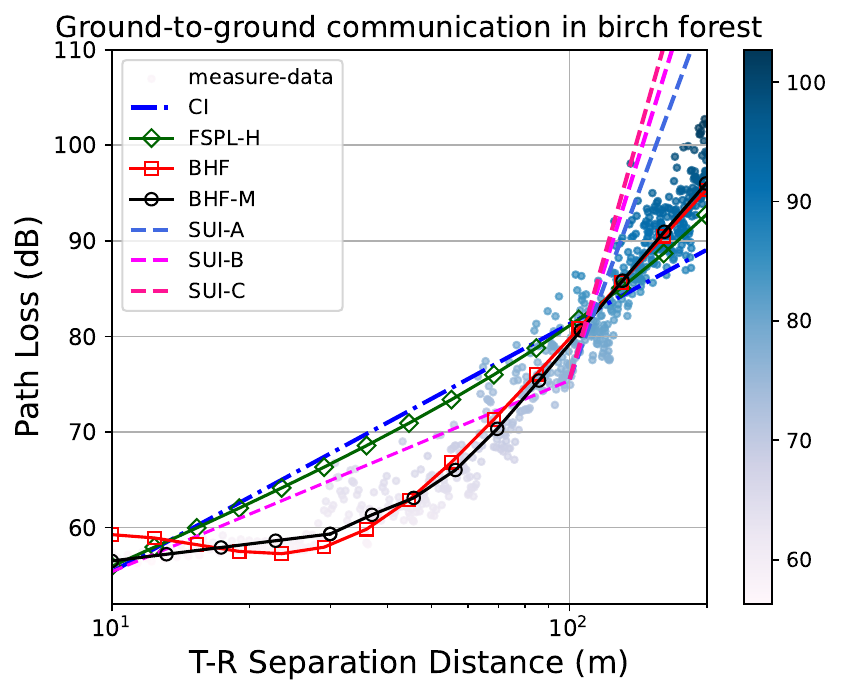}}
        \centerline{(b)}
	\end{minipage}
        
    \caption{Measured path loss data and fitting results of the close-in model, the ITU forest excess attenuation model, the SUI models, the BHF model, and the proposed BHF-M model.}
	\label{}
\end{figure}

Fig. 8 shows the fitting results of the CI, FSPL-H, BHF, SUI, and BHF-M models. It can be observed from Fig. 8 that the CI model is a straight line, because the relationship between the parameter term and the 3D T-R separation distance in the expression is multiplication, the function between path loss and distance in the logarithmic scale is linear when other coefficients are given. The three straight lines of the SUI models appear to be higher than the actual measurement data. Because the SUI model is fitted based on data from forest regions in North America, its effectiveness in fitting data from other areas could be better. The FSPL-H, BHF, and BHF-M models proposed in this paper have better fitting effects on the measured data in forest areas. As shown in Fig. 8(b), the curve of the BHF model at distances less than 30 m slowly decreases with increasing distance due to overfitting caused by the BHF model's attempt to fit data at longer distances. Therefore, the BHF-M model has been modified based on this, making the overfitting phenomenon no longer apparent and more in line with the variation law of forest path loss. We use the least squares method to determine the optimal parameters of the above model and display them in Table \uppercase\expandafter{\romannumeral2}.

\begin{table}[htbp]
\centering
\caption{PATH LOSS PARAMETERS AND PERFORMANCE FOR G2G CHANNEL}
\resizebox{1.0\linewidth}{!}{
\begin{tabular}{|c|c|c|c|c|c|}
\hline
\textbf{Env.}                 & \textbf{Model}  & \textbf{$\sigma$(dB)} & \begin{tabular}[c]{@{}l@{}}$n$(CI)\\$Am$(FSPL-H)\\$\alpha$(BHF) \\ $\alpha$\textbackslash$n$(BHF-M) \\ $a$(SUI)\end{tabular}                       &\begin{tabular}[c]{@{}l@{}}$\mu$(FSPL-H) \\$\beta$(BHF)\\$\beta$\textbackslash$m$(BHF-M) \\ $b$(SUI)\end{tabular}        &\begin{tabular}[c]{@{}l@{}}\\ $\zeta$(BHF)\\$\zeta$(BHF-M) \\ $c$(SUI)\end{tabular}        \\ \hline
\multirow{7}{*}{\textbf{Larch}}  & CI     & 5.2  & 2.6                    & -       & -      \\ \cline{2-6} 
                       & FSPL-H & 4.5  & 30                    & 0.1    & -      \\ \cline{2-6} 
                       & BHF    & 4.0  & 4.3                   & 89.0     & -42      \\ \cline{2-6} 
                       & \textbf{BHF-M}  & \textbf{3.8}  & 1.1\textbackslash{}4.3 & 33.8\textbackslash{}1.0 & -11.7 \\ \cline{2-6} 
                       & SUI-A  & 9.0 & 4.6                      & 0.0075       & 12.6      \\ \cline{2-6} 
                       & SUI-B  & 12.1 & 4                      & 0.0065       & 0.005      \\ \cline{2-6} 
                       & SUI-C  & 14.1 & 3.6                      & 0.005       & 20      \\ \hline
\multirow{7}{*}{\textbf{Birch}} & CI     & 6.0  & 2.6                    & -       & -      \\ \cline{2-6} 
                       & FSPL-H & 4.8  & 1335.0                    & 0.1    & -   \\ \cline{2-6} 
                       & BHF    & 2.8  & 5.2                   & 98.9     & -58.5      \\ \cline{2-6} 
                       & \textbf{BHF-M}  & \textbf{2.6}  & 0.6\textbackslash{}5.2 & 33.5\textbackslash{}1.0    & -14.3   \\ \cline{2-6} 
                       & SUI-A  & 8.3  & 4.6                      & 0.0075       & 12.6      \\ \cline{2-6} 
                       & SUI-B  & 11.3 & 4                      & 0.0065       & 17.1      \\ \cline{2-6} 
                       & SUI-C  & 13.4 & 3.6                      & 0.005       & 20      \\ \hline
\end{tabular}}
\end{table}

Table \uppercase\expandafter{\romannumeral2} shows the fitting error of the model to real data, and we use root mean square error (RMSE) to evaluate the model's fitting ability. It can be observed from Table \uppercase\expandafter{\romannumeral2} that due to differences in vegetation type and density, as well as the impact of geographical environment on signal transmission, the fitting error of the birch forest is generally smaller than that of the larch forest. This is because birch belongs to a broad-leaved forest while larch belongs to a coniferous forest, and birch trees' distribution is more uniform than larch. The RMSE of the models proposed in this paper in birch and larch forests are 4.9 dB and 2.6 dB, respectively. Among all models, the fitting error is the smallest, and the reduction effect on the fitting error of path loss in birch forests is significant. The RMSE calculated in the present invention is 46$\%$ lower than that of the best-performing Okumura-Hata model in other traditional models, indicating that the overall fitting effect of the wireless propagation path loss model proposed in this paper is the best, more suitable for predicting path loss in forest scenes. Compared to the BHF model, the BHF-M model decreased by 6$\%$ in larch forests and 19$\%$ in birch forests, indicating that the modified model performs better in fitting forest scene losses.

\subsection{A2G path loss model}

We measured channel data for three elevation angles in two forest environments for A2G channel measurement and used CI, ITU-slant, and Okumura-Hata models to fit channel data. The excess path loss caused by forest expression of the ITU slant forest model is as follows:

\begin{small}
\begin{equation}
\mathrm{P L}_{\mathrm{ITU-S}}=A f^{B} d_v^{C}(\theta+E)^{G}
\end{equation}
\end{small} 

\noindent where $f$ is the frequency in MHz, $d_v$ is the vegetation depth in meters, and $\theta$ is the elevation angle in degrees. A, B, C, E, and G are parameters. The expression for the entire path loss model FSPL-S is as follows:

\begin{equation}\text{PL}_{\text{FSPL-S}}=20\log_{10}(\frac{4\pi\textit{fd}\times10^9}{c})+Af^B{d_v}^C{(\theta+E)}^G,\end{equation}

\noindent where $d$ is the 3-D T-R separation distance in kilometers.

The base station antenna height of the Okumura-Hata model should be greater than 30 m, and the height of the UAV meets this condition for A2G channels.

Observing the measured data, we found that the fluctuation of path loss is relatively large compared to the G2G channel, which may be due to the higher energy of the signal component generated by ground reflection in the A2G channel. Therefore, we attempted to use a two-path model to represent this process. Due to the short measurement distance of our channel, the results of the flat Earth model and the curved Earth model are almost identical, so we use the flat Earth two-ray model (FE2R) \cite{37}. In \cite{28}, the authors employed the FE2R model to characterize the channel between the air and the sea surface. In \cite{hill}, the authors utilized the FE2R model to represent the A2G channel in mountainous environments. Fig. 9 shows the geometry of the FE2R model and the expression of the FE2R model as follows
\begin{figure}[htbp]
\centerline{\includegraphics[width=\linewidth]{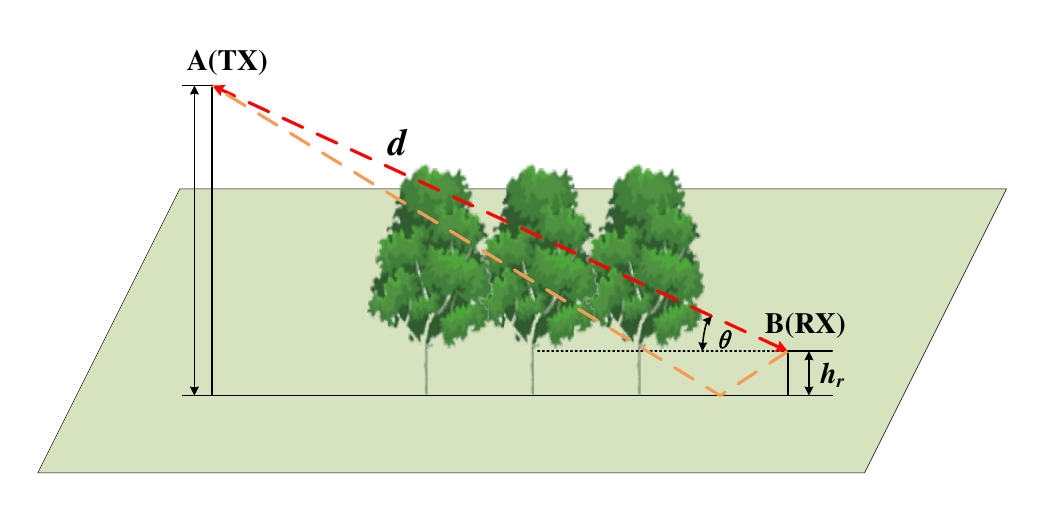}}
\caption{Geometry for FE2R model.}
\label{fig}
\end{figure}

\begin{subequations}
\begin{equation}
\mathrm{P L}_{\mathrm{FE2R}}=-20\lg[\frac\lambda{4\pi}(\frac1d+\frac{R\operatorname{e}^{-j\Delta\varphi}}{d’})]
,\end{equation}
\begin{equation}R=\frac{\sin\theta-z}{\sin\theta+z},z=\sqrt{\xi_r-\cos^2\theta/\xi_r}
,\end{equation}
\begin{equation}
\Delta\boldsymbol{\varphi}=\frac{2\pi(d'-d)}\lambda  
,\end{equation}
\begin{equation}
d'=\sqrt{(d\cos\theta)^2+\left(d\sin\theta+2h_r\right)^2}
,\end{equation}
\end{subequations}
\noindent where $d$ is the distance of the direct component between the TX and the RX, $d'$ is the distance of the reflected component from the ground to the RX, $R$ is the ground reflection coefficient,$\xi_r$ is the dielectric constant of the earth, and after considering the real situation and consulting relevant literature, we chose an average value of 15 \cite{37}.  $\theta$ is the elevation angle. $\Delta\boldsymbol{\varphi}$ is the phase difference between the direct and reflected signals. $h_r$ is the height of the RX. By observing CIR in Fig. 12, it can be seen that in addition to these two paths, there are many other multipath signals in the A2G channel of the forest area. We have divided the signal components into three parts: the LoS path, the LoS cluster, and other scattered signal energies. The total energy of other scattered signals is very low and can therefore be neglected. In the LoS cluster, the energy of ground reflection signals is relatively high, while other signals within the cluster are independent random signals. Therefore, when calculating the path loss, we used the FE2R model as the foundation and to improve it, resulting in the FE2R-M model. The modified FE2R-M model is as follows:
\begin{subequations}
\begin{equation}
\mathrm{P L}_{\mathrm{FE2R-M}}=-20n\lg[\frac\lambda{4\pi}(\frac1d+\frac{R\operatorname{e}^{-j\Delta\varphi}}{d’})]+m 
,\end{equation}

\begin{equation}
d'=\sqrt{(d\cos\theta)^2+\left(d\sin\theta+2h_r\right)^2}+l
,\end{equation}
\end{subequations}

\begin{figure*}[htbp]
\centering
	\begin{minipage}{0.48\linewidth}
		\centerline{\includegraphics[width=\textwidth]{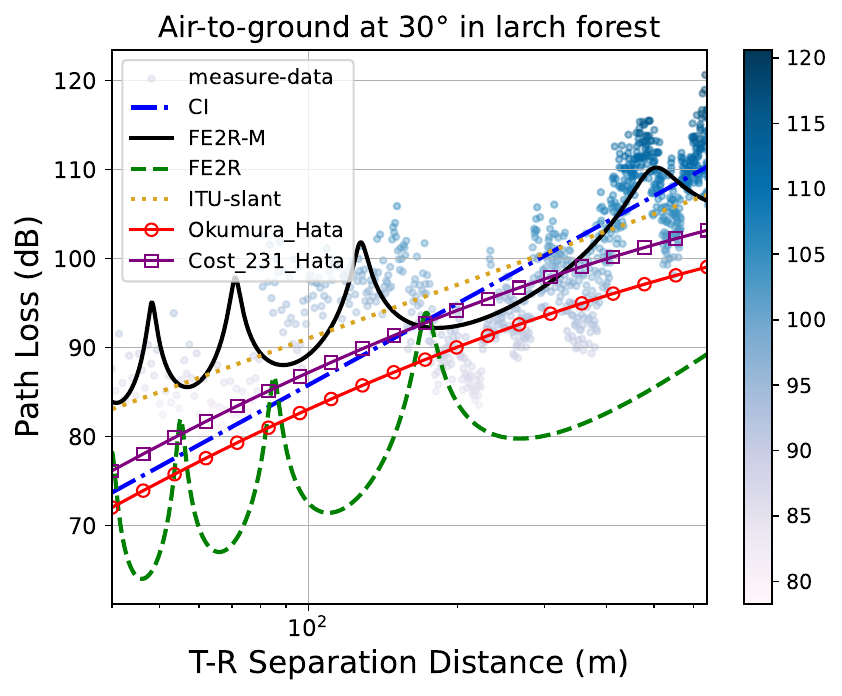}}
	 \centerline{(a)}
	\end{minipage}
	\begin{minipage}{0.48\linewidth}
		\centerline{\includegraphics[width=\textwidth]{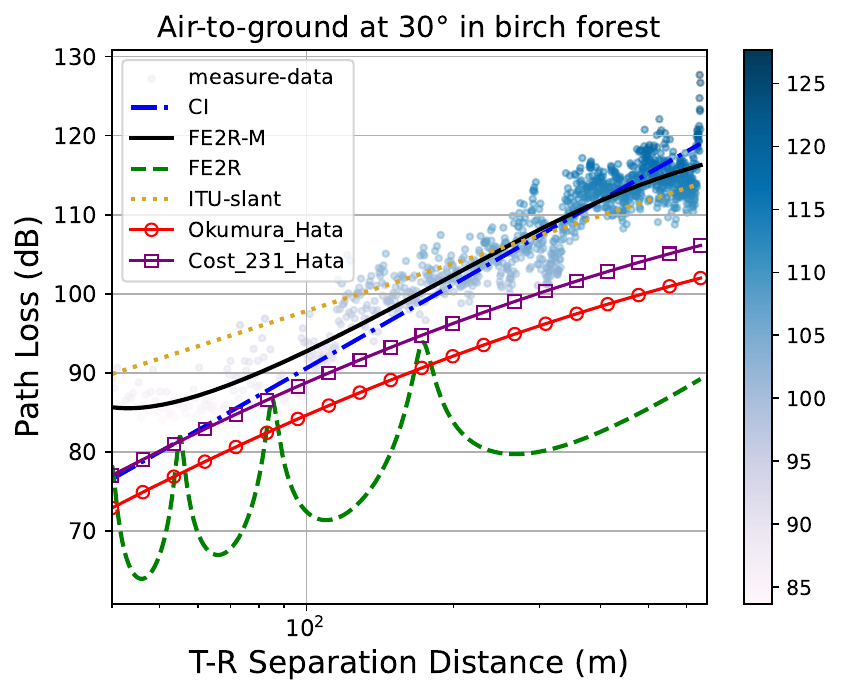}}
        \centerline{(b)}
	\end{minipage} 
    \begin{minipage}{0.48\linewidth}
		\centerline{\includegraphics[width=\textwidth]{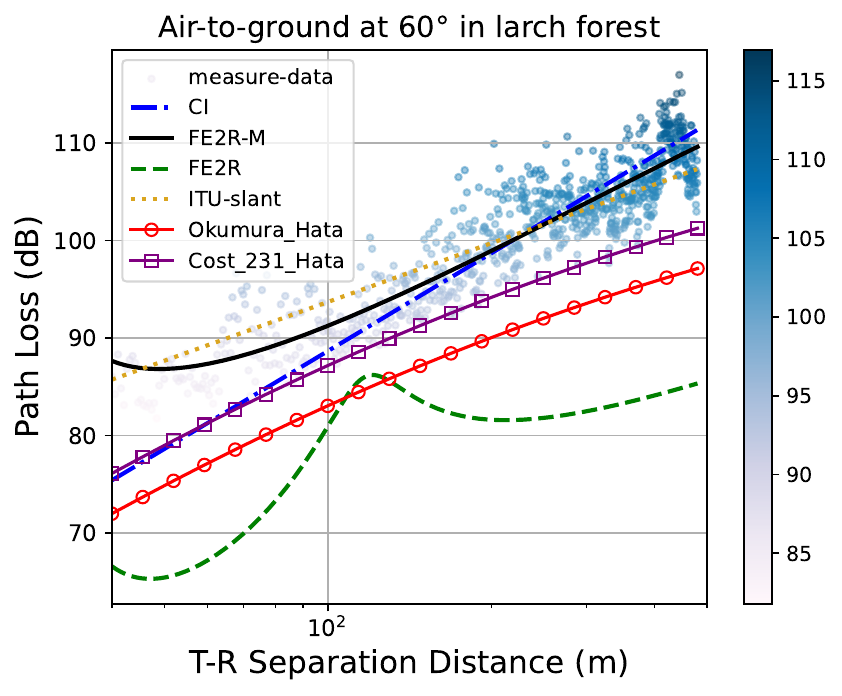}}
	 \centerline{(c)}
	\end{minipage}
 \begin{minipage}{0.48\linewidth}
		\centerline{\includegraphics[width=\textwidth]{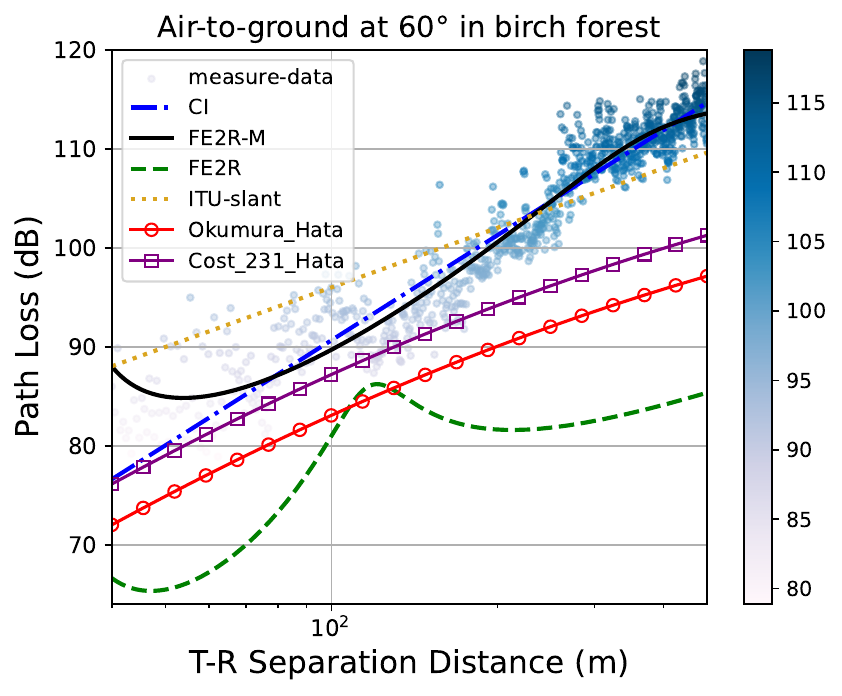}}
	 \centerline{(d)}
	\end{minipage}
 \begin{minipage}{0.48\linewidth}
		\centerline{\includegraphics[width=\textwidth]{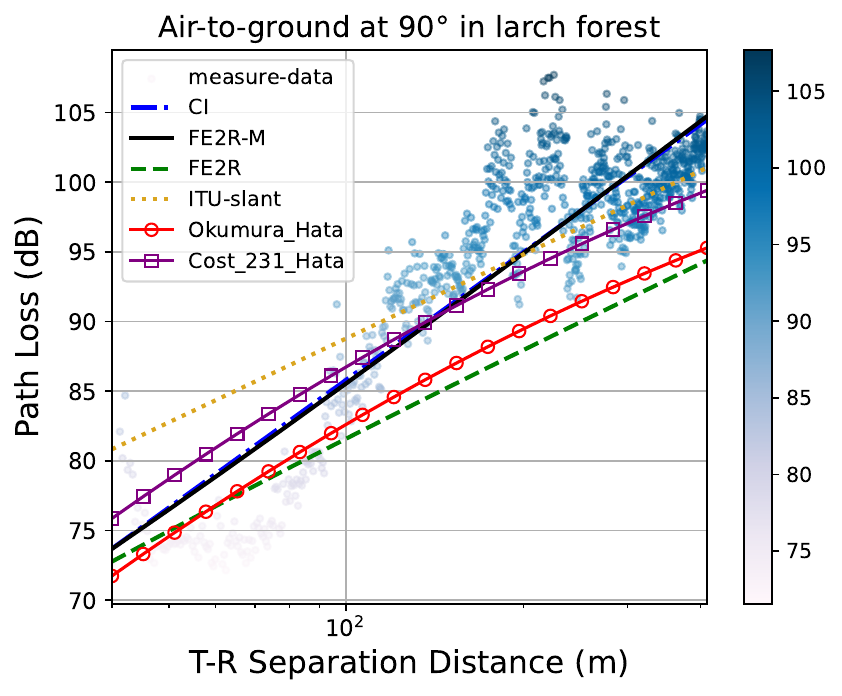}}
	 \centerline{(e)}
	\end{minipage}
 \begin{minipage}{0.48\linewidth}
		\centerline{\includegraphics[width=\textwidth]{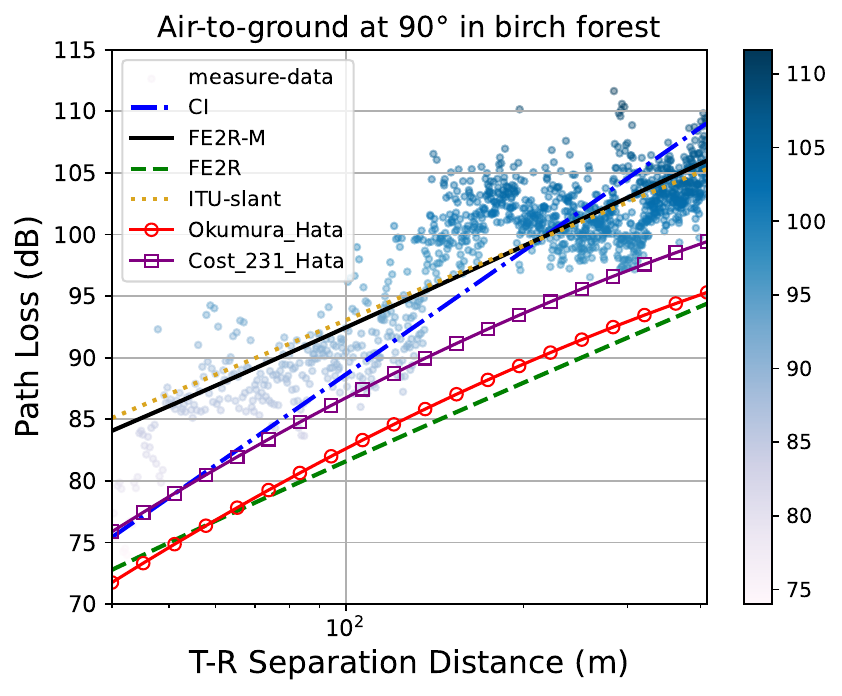}}
	 \centerline{(f)}
	\end{minipage}
    \caption{The measured path loss data and fitting results, including two forest environments, three elevation angles, and six resulting figures: (a) A2G at 30° in larch forest. (b) A2G at 30° in birch forest. (c) A2G at 60° in larch forest. (d) A2G at 60° in birch forest. (e) A2G at 90° in larch forest. (f) A2G at 90° in birch forest.}
	\label{}
\end{figure*}

\noindent where $n$ denotes the path loss exponent, representing the degree of attenuation of environmental impact signals; $l$ represents the optimized bias of the signal passing through vegetation and reflecting through the ground; $m$ represents the energy obtained by superimposing other signals within the LoS cluster in the forest area. These modifications allow the model to capture the impact of forested terrains on signal propagation more accurately.

Due to the limitation of flight altitude, we maintain the UAV at a maximum altitude of less than 500 m, usually around 400 m, during measurement. At the same time, the surrounding environment can also affect the drone's flight path. Therefore, the maximum distance we measure at different angles may differ slightly. The distance between the TX and RX at an elevation angle of 30° is around 640 m, the distance between the TX and RX at an elevation angle of 60° is about 480 m, and the distance between the TX and RX at an elevation angle of 90° is around 410 m.  Fig. 10 shows the fitting results of six path loss curves for three different elevation angles in two forest environments, where Fig (a), (c), and (e) are in a larch forest and Fig (b), (d), and (f) are in a birch forest.  The results show that the oscillation degree of FE2R curves varies with different elevation angles. The curve with an elevation angle of 30° has four waves, with the largest amplitude of oscillation, and the curve with an elevation angle of 60° has one wave, while the curve with an elevation angle of 90° is close to a straight line without any waves. The main reason is that the distance difference between the two paths in the FE2R model is related to the degree of elevation angle. Different distance differences will result in phase differences of other signals, leading to varying results after the superposition of the two signals.
Table \uppercase\expandafter{\romannumeral3} shows the fitting results of different channel models and provides the fitting parameters for these models. From Table \uppercase\expandafter{\romannumeral3}, it can be seen that due to differences in vegetation type and density, as well as the influence of the geographical environment on signal transmission, the fitting error for the birch forest is usually smaller than that for the larch forest, which is consistent with the results for the G2G channel mentioned earlier. Additionally, the fitting results of the FE2R-M model proposed in this article are the best across all scenarios and are more suitable for application in air-ground communication in forest environments.


\begin{table*}[htbp]
\caption{PATH LOSS PARAMETERS AND PERFORMANCE FOR A2G CHANNEL}
\resizebox{1.0\linewidth}{!}{
\begin{tabular}{|c|c|c|c|c|c|c|c|c|c|}
\hline
\multirow{3}{*}{\textbf{Env.}}                                                    & \multirow{3}{*}{\textbf{Ang.}} & \multirow{3}{*}{\textbf{Model}} & $n$(CI)   &     &      &     & \multirow{3}{*}{$\sigma$(dB)} & \multirow{3}{*}{\textbf{Model}} & \multirow{3}{*}{$\sigma$(dB)} \\
                                                                         &                       &                        & $A$(FSPL-S)  & $B$(FSPL-S)    & $C$(FSPL-S)     & $G$(FSPL-S)    &                       &                        &                       \\
                                                                         &                       &                        & $n$(FE2R-M)    & $m$(FE2R-M)    & $l$(FE2R-M)     &     &                       &                        &                       \\ \hline
\multirow{9}{*}{\begin{tabular}[c]{@{}c@{}}\textbf{Larch} \end{tabular}} & \multirow{3}{*}{30}                    & CI                     & 2.8 & -    & -     & -    & 6.4                   & Okumura\_Hata          & 5.9                   \\ \cline{3-10} 
                                                                         &                       & FSPL-S              & 0.2 & 0.4  & 0.2   & 0.1    & 5.9                   & Cost\_231\_Hata        &6.4                    \\ \cline{3-10} 
                                                                         &                       & \textbf{FE2R-M}                & 1.0 & 0.6 & 45.6  & -    & \textbf{4.9}                   & FE2R                   & 16.5                  \\ \cline{2-10} 
                                                                         & \multirow{3}{*}{60}   & CI                     & 3.3 & -    & -     & -    & 3.7                   & Okumura\_Hata          & 10.6                  \\ \cline{3-10} 
                                                                         &                       & FSPL-S              & 0.4 & 0.3  & 0.4  & 0.0  & 3.3                   & Cost\_231\_Hata        & 6.8                  \\ \cline{3-10} 
                                                                         &                       & \textbf{FE2R-M}                & 0.9 & 0.7  & 37.9  & -    & \textbf{2.9}                   & FE2R                   & 19.8                  \\ \cline{2-10} 
                                                                         & \multirow{3}{*}{90}   & CI                     & 3.0 & -    & -     & -    & 3.8                   & Okumura\_Hata          & 7.2                   \\ \cline{3-10} 
                                                                         &                       & FSPL-S              & 0.4 & 0.1  & 0.4   & 0.3 & 4.7                   & Cost\_231\_Hata        & 4.6                   \\ \cline{3-10} 
                                                                         &                       & \textbf{FE2R-M}                & 1.1 & 0.9  & 11.6 & -    & \textbf{3.5}                   & FE2R                   & 8.3                   \\ \hline
\multirow{9}{*}{\begin{tabular}[c]{@{}c@{}}\textbf{Birch} \end{tabular}}  & \multirow{3}{*}{30}   & CI                     & 3.5 & -    & -     & -    & 3.4                   & Okumura\_Hata          & 12.4                  \\ \cline{3-10} 
                                                                         &                       & FSPL-S              & 0.3 & 0.3  & 0.3   & 0.3  & 4.0                   & Cost\_231\_Hata        & 8.6                   \\ \cline{3-10} 
                                                                         &                       & \textbf{FE2R-M}                & 1.0 & 0.8  & 29.7 & -    & \textbf{2.8}                  & FE2R                   & 25.6                  \\ \cline{2-10} 
                                                                         & \multirow{3}{*}{60}   & CI                     & 3.5 & -    & -     & -    & 3.0                   & Okumura\_Hata          & 13.2                  \\ \cline{3-10} 
                                                                         &                       & FSPL-S              & 0.6 & 0.3  & 0.0  & 0.4 & 4.7                   & Cost\_231\_Hata        & 9.4                   \\ \cline{3-10} 
                                                                         &                       & \textbf{FE2R-M}                & 0.9 & 0.9  & 25.2  & -    & \textbf{2.6}                   & FE2R                   & 22.5                  \\ \cline{2-10} 
                                                                         & \multirow{3}{*}{90}   & CI                     & 3.2 & -    & -     & -    & 3.4                   & Okumura\_Hata          & 9.0                   \\ \cline{3-10} 
                                                                         &                       & FSPL-S              & 0.7 & 0.1  & 1.3   & -0.4 & 3.7                   & Cost\_231\_Hata        & 5.3                   \\ \cline{3-10} 
                                                                         &                       & \textbf{FE2R-M}                & 1.1 & 1.4  & -14.9 & -    & \textbf{3.1}                   & FE2R                   & 10.1                  \\ \hline
\end{tabular}}
\end{table*}

\subsection{Shadow Fading}

During signal propagation, obstacles such as forests and shrubs can obstruct the signal, leading to the formation of shadowed regions in the receiving area. This phenomenon is referred to as shadow fading. As illustrated in Fig. 10, the path loss exhibits noticeable oscillations, which is a direct consequence of shadow fading. Table \uppercase\expandafter{\romannumeral5} further supports this observation, showing that the variance of shadow fading is highest for Fig. 10 (a), indicating more significant signal fluctuations in this scenario. The measurement environment corresponding to Fig. 10 (a) is a larch forest, characterized by trees with larger trunk diameters and greater heights compared to those in birch forests. Although the tree density in the larch forest is relatively lower, the larger dimensions of individual trees result in stronger signal obstruction and more pronounced shadow fading. Additionally, the measurement geometry, including the relative positioning of the TX and RX, contributes to the observed oscillations.
Fig. 11 presents the probability density function (PDF) of shadowing variables in both larch and birch forest environments. The results reveal that the variance of path loss in birch forests is greater than that in larch forests, a distinction attributed to the differences in vegetation types. Table \uppercase\expandafter{\romannumeral5} provides a detailed description of the magnitude of shadowing variables for both A2G and G2G signals in these two environments.

\begin{figure}[t]
\centerline{\includegraphics[width=\linewidth]{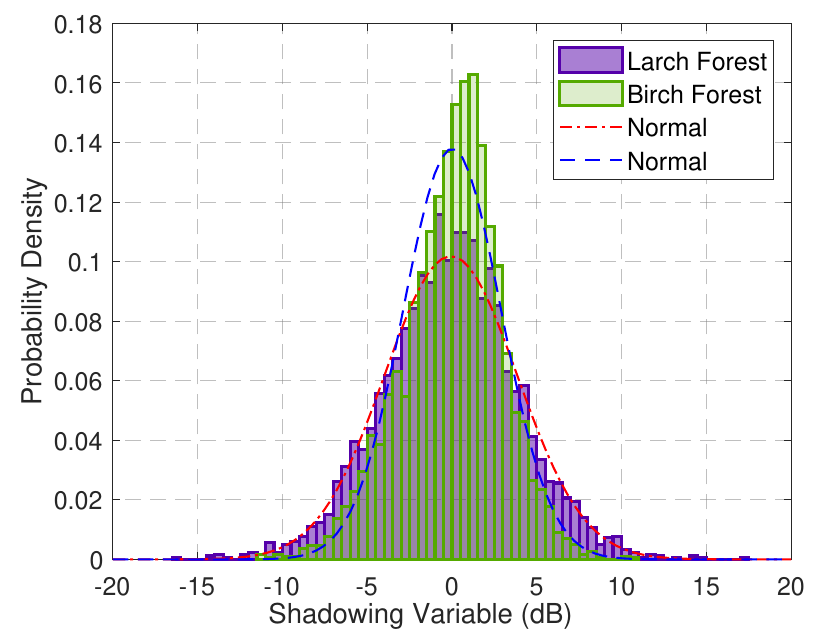}}
\caption{PDF of shadowing variable in larch forests and birch forests.}
\label{fig}
\end{figure}

\section{Spatio-Temporal Statistics}
This section will study small-scale channel parameters in the forest environment. This section consists of three parts: extracting CIR, calculating the number of multipaths, and extracting small-scale channel parameters.
\subsection{Channel Impulse Response}

We collected OFDM signal data at the RX through on-site measurements, stored them in hexadecimal format, and divided them into I/Q signals. By combining these signals through binary conversion, we obtain the time-domain signal at the RX. The Zadoff-Chu (ZC) sequence, widely used as a constant amplitude zero auto-correlation (CAZAC) sequence, offers excellent correlation properties. Additionally, \cite{38} proposed joint fine timing and channel estimation using ZC or general CAZAC sequences for downlink transmission, which leverages cyclic suffixes to estimate the channel impulse response via cross-correlation with the received ZC sequence. This approach provides a computationally efficient and accurate channel impulse response estimation method. Fig. 12 depicts the derived channel impulse response waveform obtained through cross-correlation between the received and template signals.  We also obtained the channel frequency response (CFR) based on the OFDM signal frame structure information in section \uppercase\expandafter{\romannumeral2}. We verified that the CIR obtained by correlating the IFFT of CFR with the ZC sequence is consistent. Fig. 12 (a) shows the CIR for A2G and G2G channels. Fig. 12 (b) shows the real parts of the CFR for A2G and G2G channels.

\begin{figure*}[h]
	\begin{minipage}{0.5\linewidth}
		\centerline{\includegraphics[width=\textwidth]{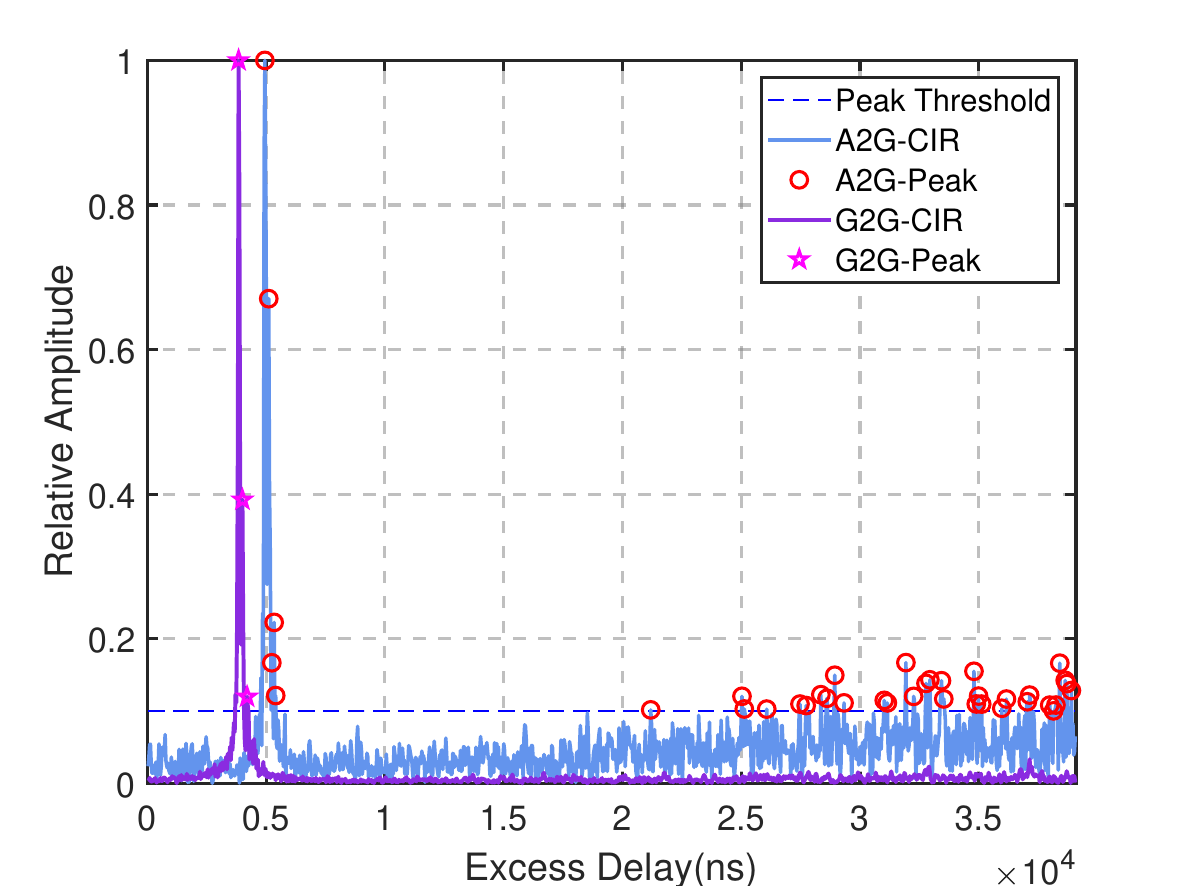}}
	 \centerline{(a)}
	\end{minipage}
	\begin{minipage}{0.5\linewidth}
		\centerline{\includegraphics[width=\textwidth]{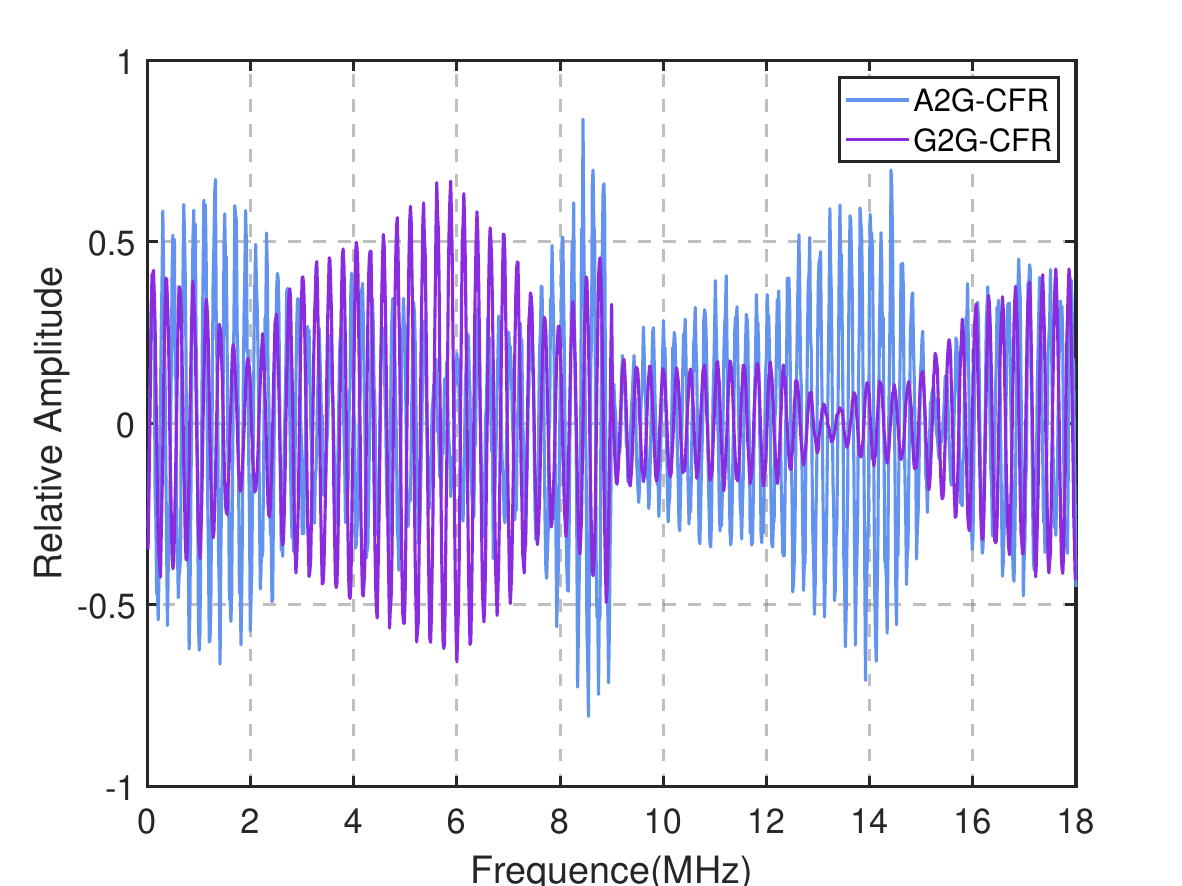}}
        \centerline{(b)}
	\end{minipage} 
        
    \caption{CIR and CFR in A2G and G2G channels. (a) CIR. (b) CFR}
	\label{}
\end{figure*}

\begin{figure*}[h]

	\begin{minipage}{0.5\linewidth}
		\centerline{\includegraphics[width=\textwidth]{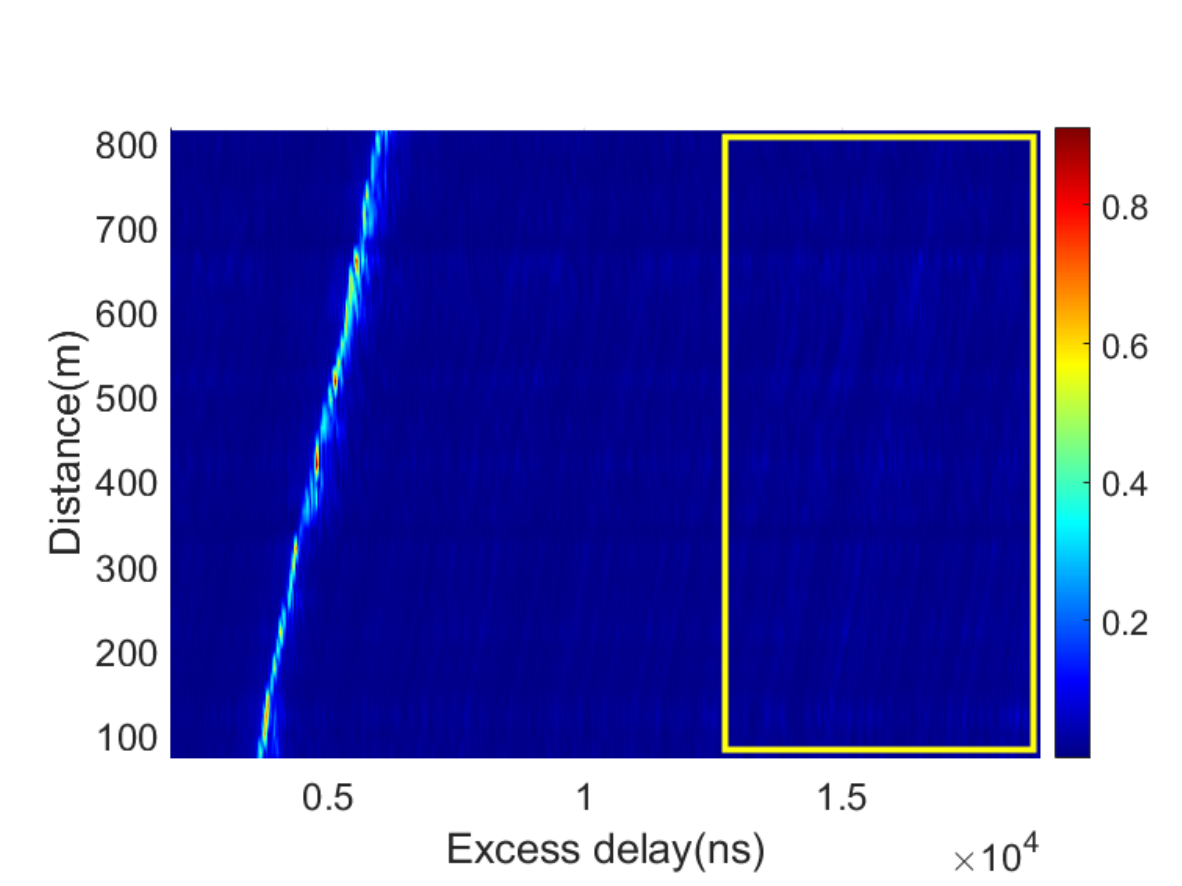}}
        \centerline{(a)}
	\end{minipage} 
    \begin{minipage}{0.5\linewidth}
        \centerline{\includegraphics[width=\textwidth]{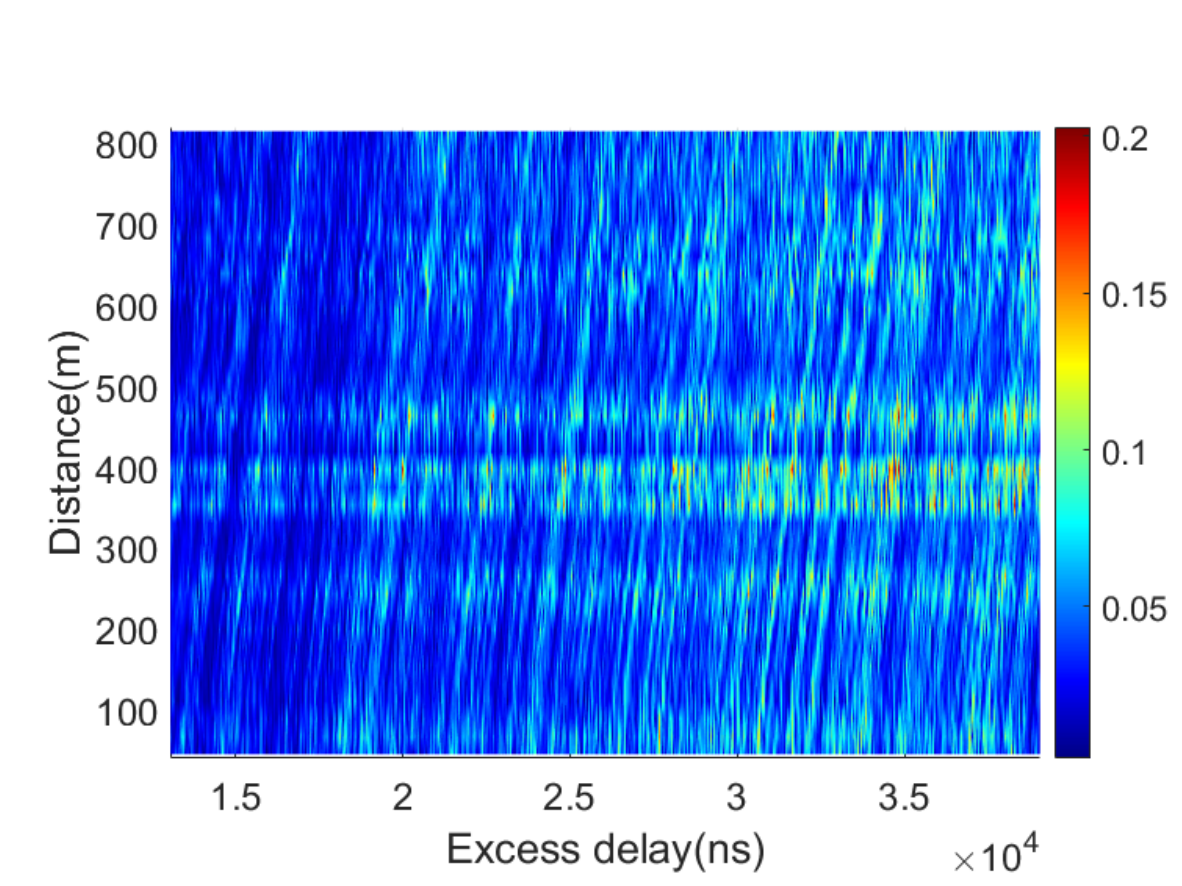}}
        \centerline{(b)}
	\end{minipage}
        
    \caption{The propagation of multipath signals in the A2G channel in larch forests. (a) vertical view of PDP. (b) Zoom-in of the boxed region in (a)}.
	\label{}
\end{figure*}

\subsection{Extracting Multipath Information}

Received signals in OFDM systems $r(t)$ and CIR of wireless channels $h(t)$ are denoted by (3) and (4), respectively. The CIR encompasses information for all subcarriers, representing the superposition of multiple channel paths, from which we extract the channel state information, including key parameters such as the number of multipath components, as well as the amplitude, delay, and phase of each path. After we filter and denoise the CIR, we perform a spectral peak search to select possible multipath signals. For determining a multipath signal, the following conditions should be met:

1. The time delay distance between adjacent peaks should be greater than the time resolution of the signal.

2. The amplitude of the peak should be greater than the amplitude of the noise.

3. The amplitude of the peak should not be less than the maximum peak amplitude of -20 dB.

Fig. 13 (a) shows the top view of the power spectral density map, and Fig. 13 (b) shows the top view of the power spectral density map for scattered signals. From Fig. 13 (a), it can be seen that there is a clear and bright slanted line, while other areas are relatively dim, indicating that delay increases with distance. It also suggests that multipath signals are concentrated near the first arrival path. When we zoom in on the area surrounded by the yellow lines in Fig. 13 (a), which is shown in Fig. 13 (b), we observe many slanted lines parallel to the bright lines. These diagonal lines represent scattering classifications caused by vegetation. The signal passes through the crown of trees since the TX is airborne when measuring the A2G channel. In G2G measurements, tree trunks mainly diffract the signal, resulting in a lower scattering component, which is characteristic of forest channels. However, the amplitude of the scattering component is very close to the amplitude of the noise, making it difficult to extract in practice.

In the forest environment, many scattering paths are generated in addition to the direct path through vegetation, so we divide the signal into direct and scattering clusters. We can observe the CIR and find that G2G and A2G have only one main energy cluster. Therefore, we combine the cluster delay with the relative delay within the cluster. At the same time, the A2G channel reveals many scattered signals with low power, close to noise, and without obvious clustering characteristics. Accordingly, we model the multipath channel in the forest area as (19): the LoS path, other multipath within the LoS cluster, and other low-power scattering paths.

\begin{equation}h(t)=a_0\delta(t-\tau_0)+\sum_{m=1}^Ma_m\delta(t-\tau_m)+\sum_{s=1}^Sa_s\delta(t-\tau_s)\end{equation}

Based on the measured data, we found that the scattered signal has low power, accounting for about 1.7$\%$ of the total signal power, close to the power of noise. The bit error rate (BER) of the received scattered signal is about 95$\%$. Therefore, the delay domain parameter RMS-DS can be approximated as:

\begin{equation}\tau_{mean}=\frac{\sum_{m=0}^MP(\tau_m)\tau_m}{\sum_{m=0}^MP(\tau_m)},\end{equation}

\begin{equation}\tau_{rms}=\sqrt{\frac{\sum_{m=0}^M\left(\tau_m-\tau_{mean}\right)^2P(\tau_m)}{\sum_{m=0}^MP(\tau_m)}},\end{equation}

\begin{equation}P(\tau_m)=E(a_m^2),\end{equation}

The calculation formula for the Rician K factor is:
\begin{equation}K=\frac{P(\tau_0)}{\sum_{m=1}^MP(\tau_m)}.\end{equation}

We calculated the RMS-DS and Rician K factors of G2G and A2G channels in the environments of larch and birch forests based on the above formula. Fig. 14 shows the RMS-DS and Rician K factor of the A2G channel at a 30° elevation angle in the birch forest environment, and it can be seen that these two parameters do not vary significantly with distance. We can also conclude from the graph that RMS-DS and Rician K factors show a negative correlation trend. This reflects the relationship between these two parameters and multipath effects, that is, the stronger the multipath effect, the larger the RMSDS, and the smaller the Rician K factor.


\begin{figure}[htbp]
\centerline{\includegraphics[width=\linewidth]{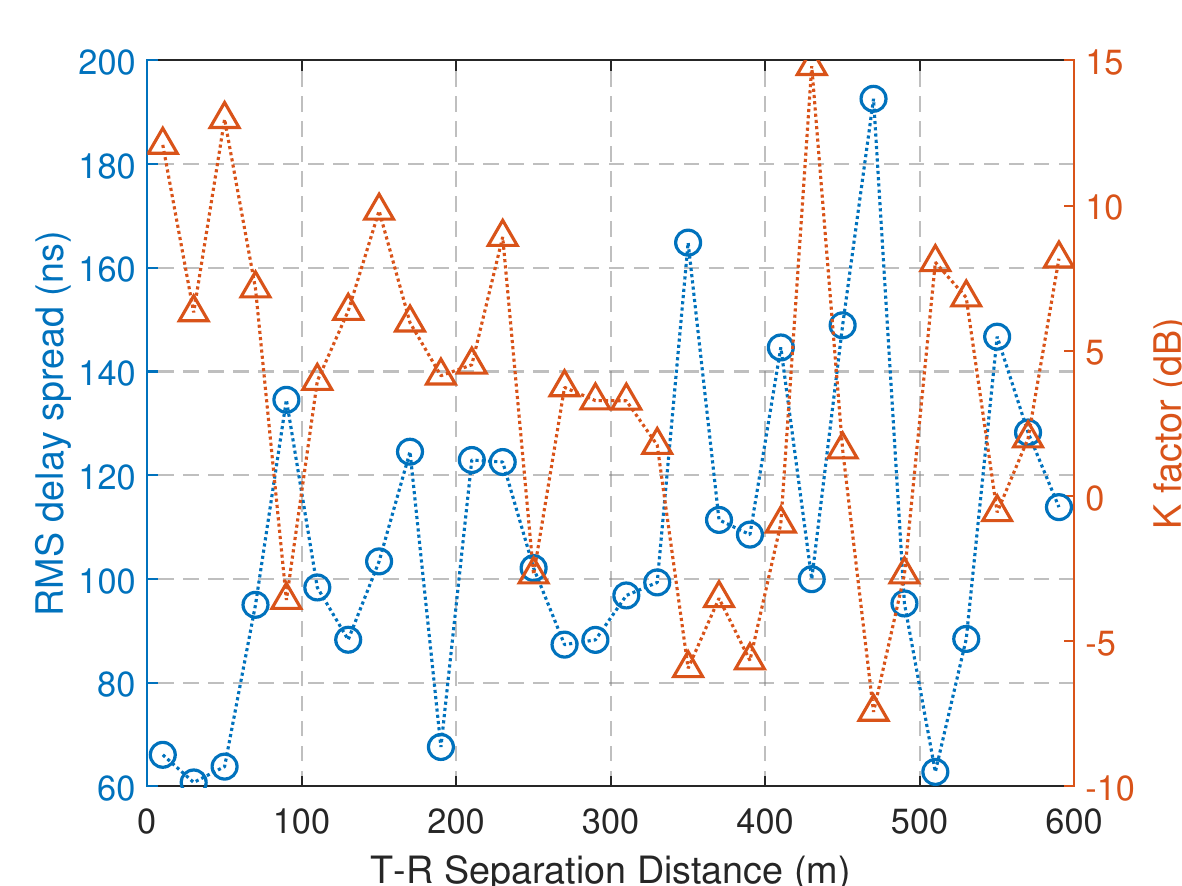}}
\caption{RMS-DS and Rician K factors in the A2G channel at a 90° elevation angle.}
\label{fig}
\end{figure}

\begin{figure}[htbp]
\centerline{\includegraphics[width=8cm]{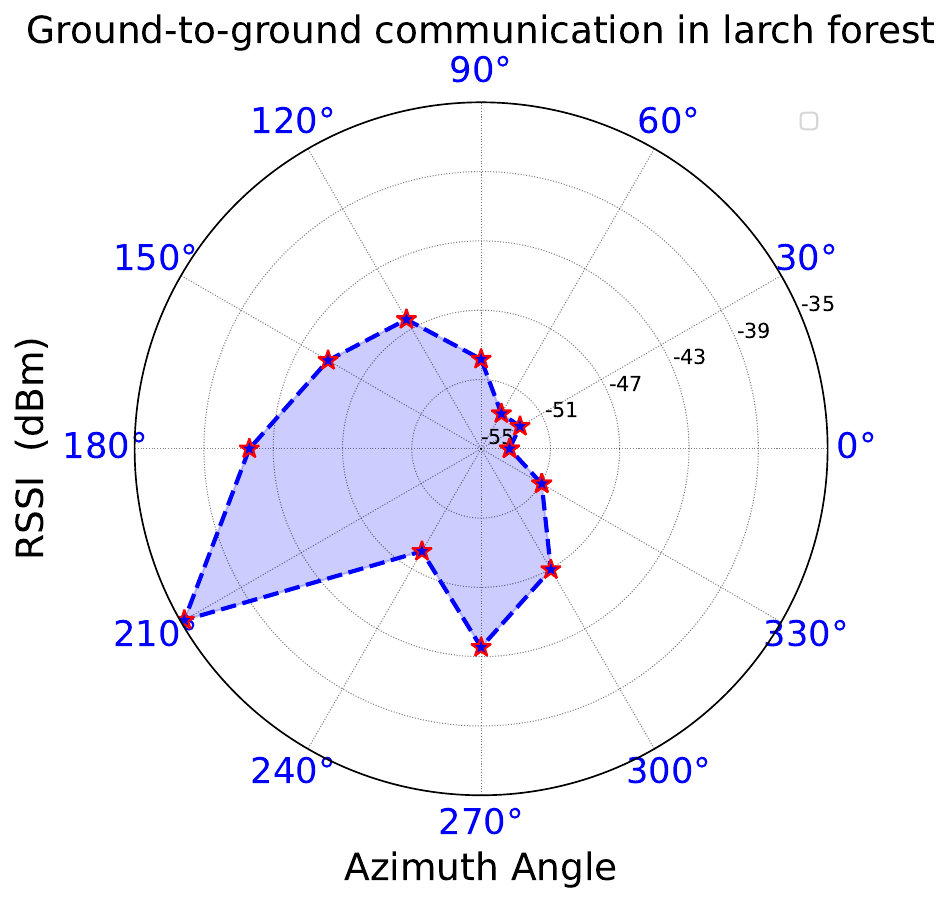}}
\caption{Autocorrelation of OFDM signals based on ZC synchronization sequence.}
\label{fig}
\end{figure}
\subsection{Angular Spread}
\begin{figure*}[h]
	
	\begin{minipage}{0.48\linewidth}
		\centerline{\includegraphics[width=\textwidth]{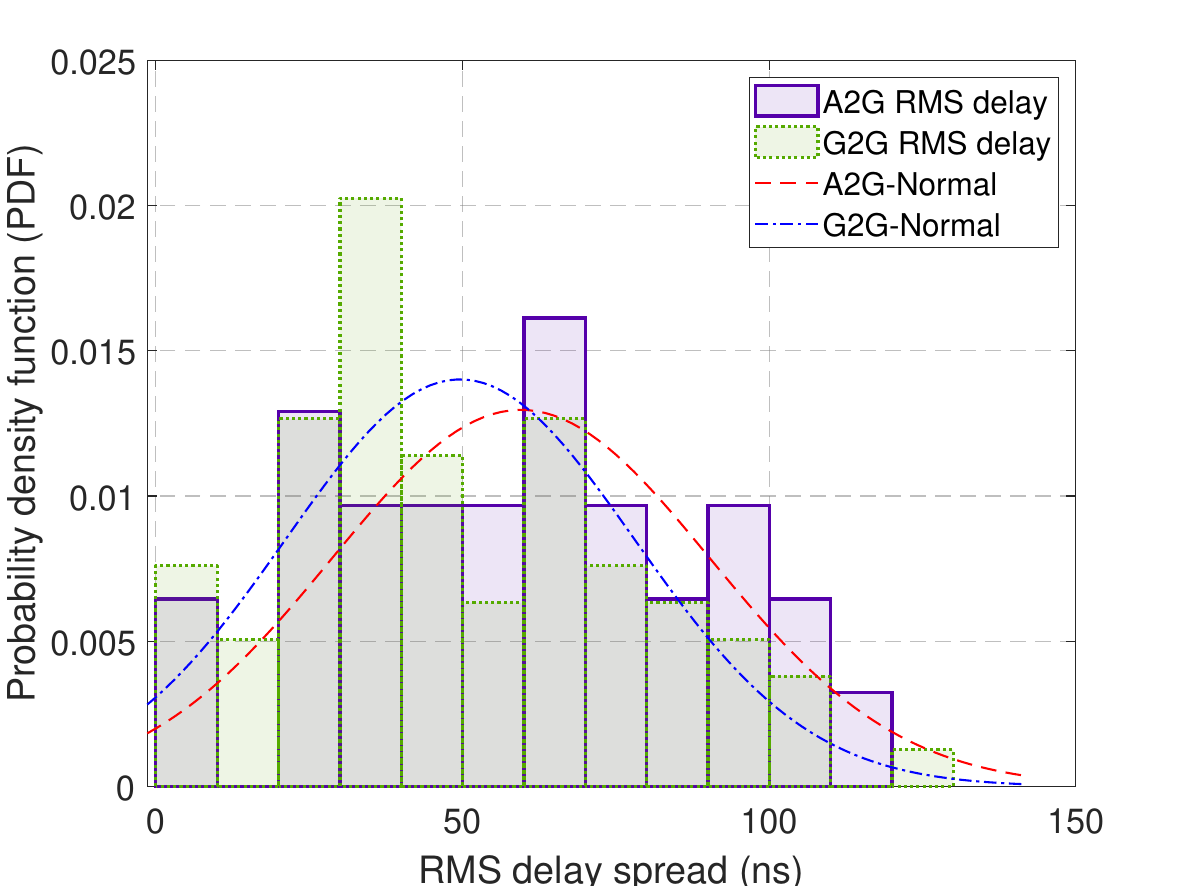}}
        \centerline{(a)}
	\end{minipage}
	\begin{minipage}{0.48\linewidth}
		\centerline{\includegraphics[width=\textwidth]{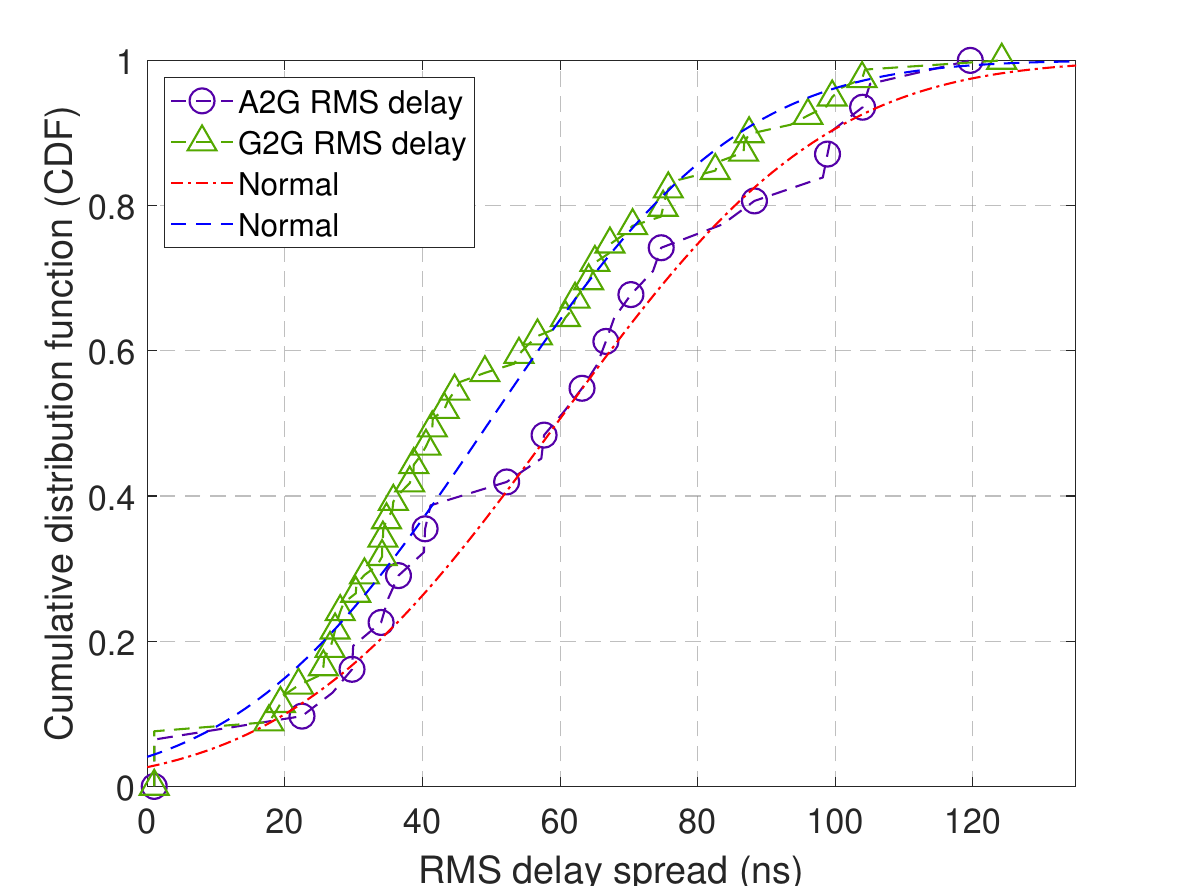}}
	 \centerline{(b)}
	\end{minipage}
        
    \caption{Distribution and fitting of RMS-DS in larch forests: (a) PDF of RMS-DS, (b) CDF of RMS-DS.}
	\label{}
\end{figure*}

\begin{figure*}[h]
	
	\begin{minipage}{0.48\linewidth}
		\centerline{\includegraphics[width=\textwidth]{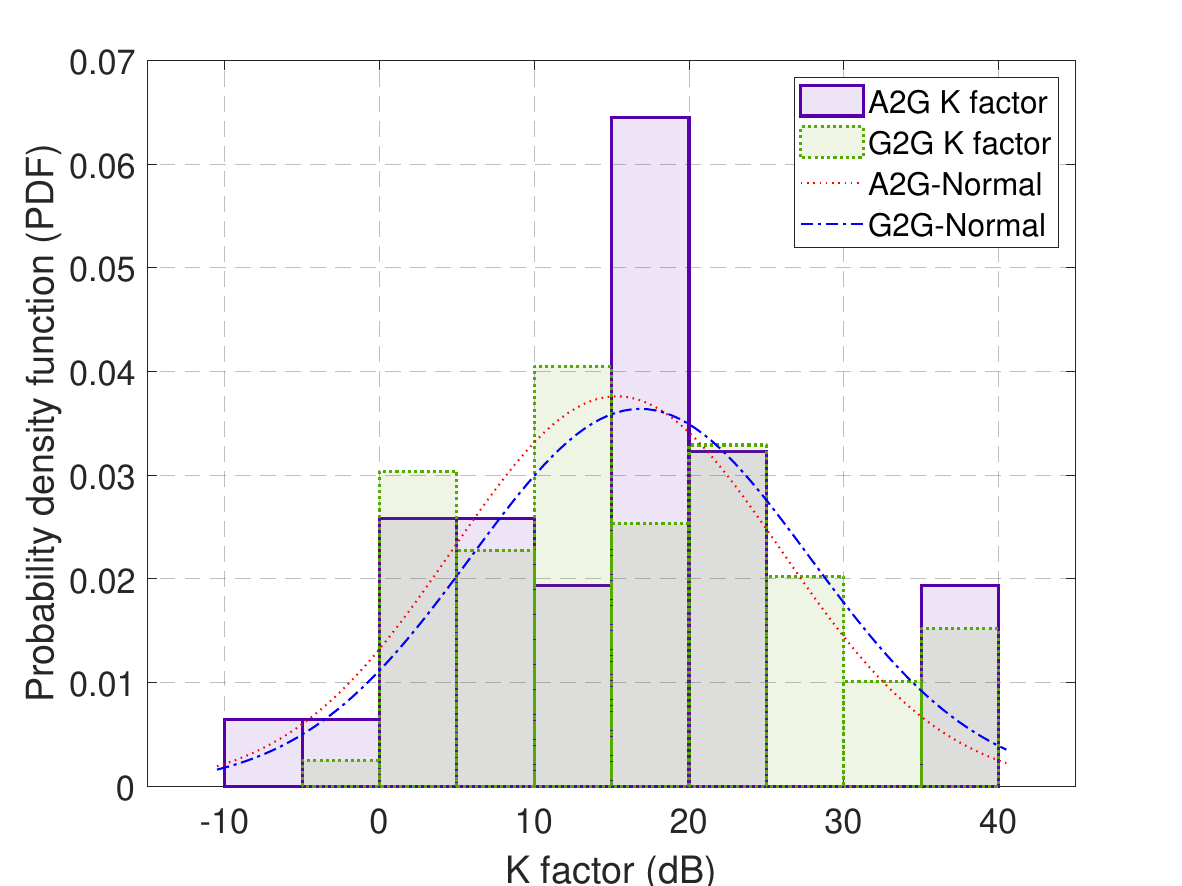}}
        \centerline{(a)}
	\end{minipage}
	\begin{minipage}{0.48\linewidth}
		\centerline{\includegraphics[width=\textwidth]{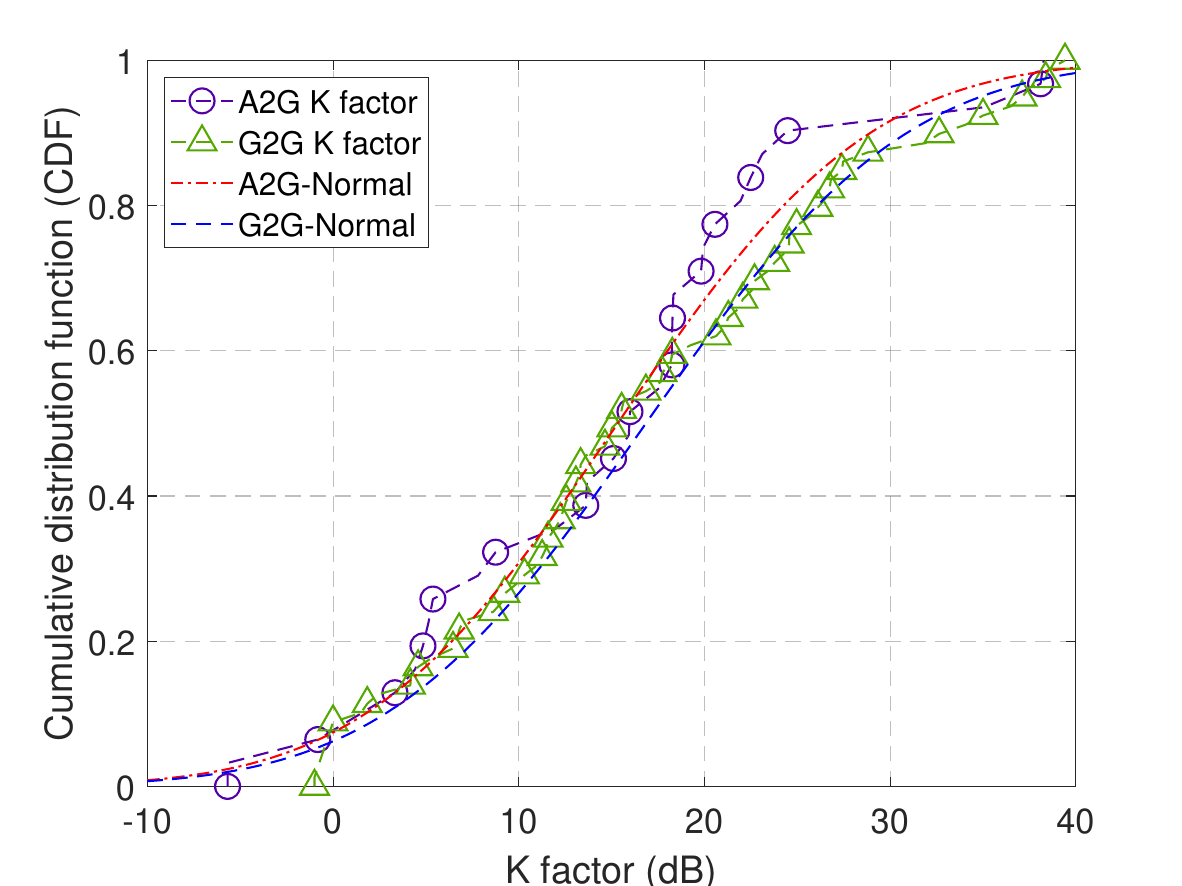}}
	 \centerline{(b)}
	\end{minipage}
        
    \caption{Distribution and fitting of Rician K factor in larch forests: (a) PDF of Rician K factor, (b) CDF of Rician K factor.}
	\label{}
\end{figure*}
APS and azimuth AS are key metrics in wireless communication systems. In our experimentation, we employed directional antennas to assess angular domain parameters across two forest environments, manipulating the azimuth angle of their half-power beamwidth. In each scenario, we collect four data sets at different distances between the TX and RX. As depicted in Fig. 15, the azimuth arrival APS was examined within a larch forest at a distance of 60 m from both the TX and RX. The results show that signals were detectable from various directions, notably with the highest signal energy observed at 210° and noticeable signal receptions at 270°. Conversely, signal energy within the 90° to 330° range experienced a decay of nearly 20 dB, indicating a minimal signal presence in this directional range.

\begin{table}[htbp]
\centering
\caption{RMS AND AVG ASA FOR EACH MEASUREMENT POSITION}
\resizebox{1.0\linewidth}{!}{
\begin{tabular}{|c|cc|cc|}
\hline
\textbf{Env.}        & \multicolumn{2}{c|}{\textbf{Larch}}                & \multicolumn{2}{c|}{\textbf{Birch}}                \\ \hline
\textbf{Dis.(m)} & \multicolumn{1}{c|}{RMS ASA(°)} & AVG ASA(°) & \multicolumn{1}{c|}{RMS ASA(°)} & AVG ASA(°) \\ \hline
10          & \multicolumn{1}{c|}{79.14}      & 150.6          & \multicolumn{1}{c|}{82.8}       & 158.7          \\ \hline
60          & \multicolumn{1}{c|}{80.2}       & 188.4          & \multicolumn{1}{c|}{114.7}      & 138.8          \\ \hline
110         & \multicolumn{1}{c|}{106.8}      & 137.1          & \multicolumn{1}{c|}{105.3}      & 129.32         \\ \hline
160         & \multicolumn{1}{c|}{106.9}      & 132.1          & \multicolumn{1}{c|}{105.4}      & 136.5          \\ \hline

\end{tabular}}
\end{table}

\begin{table*}[]
\centering
\caption{\centering{STATISTICS DISTRIBUTION PARAMETER FOR SHADOWING VARIABLE, RMS DELAY SPREAD AND RICIAN K FACTOR}}
\resizebox{1\linewidth}{!}{
\begin{tabular}{|cc|cccccccccc|}
\hline
\multicolumn{2}{|c|}{\textbf{Env.}}                                                                & \multicolumn{5}{c|}{\textbf{Larch}}                                                                                                                  & \multicolumn{5}{c|}{\textbf{Birch}}                                                                                             \\ \hline
\multicolumn{2}{|c|}{\textbf{Type}}                                                                   & \multicolumn{3}{c|}{A2G channel}                                                   & \multicolumn{1}{c|}{G2G channel} & \multicolumn{1}{c|}{Mixed} & \multicolumn{3}{c|}{A2G channel}                                                   & \multicolumn{1}{c|}{G2G channel} & Mixed \\ \hline
\multicolumn{2}{|c|}{\textbf{Ang.}}                                                                    & \multicolumn{1}{c|}{30°}    & \multicolumn{1}{c|}{60°}   & \multicolumn{1}{c|}{90°}   & \multicolumn{1}{c|}{-}           & \multicolumn{1}{c|}{-}     & \multicolumn{1}{c|}{30°}   & \multicolumn{1}{c|}{60°}   & \multicolumn{1}{c|}{90°}    & \multicolumn{1}{c|}{-}           & -     \\ \hline
\multicolumn{2}{|c|}{\textbf{Model Parameters}}                                                               & \multicolumn{10}{c|}{\textbf{Shadowing Variable Statistical Distribution(dB)}}                                                                                                                                                                                                              \\ \hline
\multicolumn{1}{|c|}{\multirow{2}{*}{Normal}}                                                & $\mu$    & \multicolumn{1}{c|}{0.0}   & \multicolumn{1}{c|}{0.0}  & \multicolumn{1}{c|}{0.0}  & \multicolumn{1}{c|}{0.0}         & \multicolumn{1}{c|}{-0.1}   & \multicolumn{1}{c|}{0.0}  & \multicolumn{1}{c|}{0.0}  & \multicolumn{1}{c|}{0.0}   & \multicolumn{1}{c|}{0.0}         & 0.0   \\ \cline{2-12} 
\multicolumn{1}{|c|}{}                                                                       & $\sigma$ & \multicolumn{1}{c|}{4.9}   & \multicolumn{1}{c|}{2.9}  & \multicolumn{1}{c|}{3.5}  & \multicolumn{1}{c|}{3.8}         & \multicolumn{1}{c|}{3.9}   & \multicolumn{1}{c|}{2.8}  & \multicolumn{1}{c|}{2.6}  & \multicolumn{1}{c|}{3.1}   & \multicolumn{1}{c|}{2.6}         & 2.9   \\ \hline
\multicolumn{2}{|c|}{FE. RMSE(\%)}                                                                    & \multicolumn{1}{c|}{0.6}    & \multicolumn{1}{c|}{1.7}   & \multicolumn{1}{c|}{1.8}   & \multicolumn{1}{c|}{2.1}           & \multicolumn{1}{c|}{1.5}     & \multicolumn{1}{c|}{0.4}   & \multicolumn{1}{c|}{1.5}   & \multicolumn{1}{c|}{1.1}    & \multicolumn{1}{c|}{1.1}           & 0.3     \\ \hline
\multicolumn{2}{|c|}{\textbf{Model Parameters}}                                                               & \multicolumn{10}{c|}{\textbf{RMS Delay Spread Statistical Distribution(ns)}}                                                                                                                                                                                                                \\ \hline
\multicolumn{1}{|c|}{\multirow{2}{*}{Normal}}                                                & $\mu$    & \multicolumn{1}{c|}{59.4} & \multicolumn{1}{c|}{51.5} & \multicolumn{1}{c|}{42.6} & \multicolumn{1}{c|}{49.5}        & \multicolumn{1}{c|}{51.0}  & \multicolumn{1}{c|}{107.6} & \multicolumn{1}{c|}{73.5} & \multicolumn{1}{c|}{73.1}  & \multicolumn{1}{c|}{80.1}        & 84.2 \\ \cline{2-12} 
\multicolumn{1}{|c|}{}                                                                       & $\sigma$ & \multicolumn{1}{c|}{30.9}  & \multicolumn{1}{c|}{16.5} & \multicolumn{1}{c|}{21.0} & \multicolumn{1}{c|}{28.6}        & \multicolumn{1}{c|}{27.1}  & \multicolumn{1}{c|}{31.4} & \multicolumn{1}{c|}{27.1} & \multicolumn{1}{c|}{25.4} & \multicolumn{1}{c|}{34.0}        & 33.2  \\ \hline
\multicolumn{2}{|c|}{FE. RMSE(\%)}                                                                    & \multicolumn{1}{c|}{0.1}    & \multicolumn{1}{c|}{0.3}   & \multicolumn{1}{c|}{0.2}   & \multicolumn{1}{c|}{0.1}           & \multicolumn{1}{c|}{0.1}     & \multicolumn{1}{c|}{0.2}   & \multicolumn{1}{c|}{0.3}   & \multicolumn{1}{c|}{0.2}    & \multicolumn{1}{c|}{0.2}           & 0.2     \\ \hline
\multicolumn{2}{|c|}{\textbf{Model Parameters}}                                                               & \multicolumn{10}{c|}{\textbf{Rician K Factor Statistical Distribution(dB)}}                                                                                                                                                                                                                 \\ \hline
\multicolumn{1}{|c|}{\multirow{2}{*}{Normal}}                                                & $\mu$    & \multicolumn{1}{c|}{14.2}   & \multicolumn{1}{c|}{15.1} & \multicolumn{1}{c|}{23.1}  & \multicolumn{1}{c|}{19.8}         & \multicolumn{1}{c|}{17.4}   & \multicolumn{1}{c|}{3.4} & \multicolumn{1}{c|}{13.1} & \multicolumn{1}{c|}{7.9}  & \multicolumn{1}{c|}{10.2}        & 7.6  \\ \cline{2-12} 
\multicolumn{1}{|c|}{}                                                                       & $\sigma$ & \multicolumn{1}{c|}{10.6}   & \multicolumn{1}{c|}{7.2} & \multicolumn{1}{c|}{11.1}  & \multicolumn{1}{c|}{11.3}         & \multicolumn{1}{c|}{10.5}   & \multicolumn{1}{c|}{5.8}  & \multicolumn{1}{c|}{10.1}  & \multicolumn{1}{c|}{8.7}  & \multicolumn{1}{c|}{9.8}        & 9.0  \\ \hline
\multicolumn{2}{|c|}{FE. RMSE(\%)}                                                                     & \multicolumn{1}{c|}{0.8}    & \multicolumn{1}{c|}{0.4}   & \multicolumn{1}{c|}{0.5}   & \multicolumn{1}{c|}{0.3}           & \multicolumn{1}{c|}{0.4}     & \multicolumn{1}{c|}{0.2}   & \multicolumn{1}{c|}{0.4}   & \multicolumn{1}{c|}{0.5}    & \multicolumn{1}{c|}{0.4}           & 0.3     \\ \hline
\end{tabular}}
\end{table*} 
Table \uppercase\expandafter{\romannumeral4} lists the RMS and AVG ASA (Azimuth Angular spread of Arrival) for each measurement position in two environments.

\subsection{RMS-DS}

RMS-DS is a crucial parameter for characterizing the properties of the delay of the wireless channel and the effects of multiple paths. Using the methodology described above, we calculated the RMS-DS values for the G2G and A2G channel within a larch forest environment. Fig. 16 shows the probability distribution of RMS-DS for G2G and A2G with an elevation angle of 30 ° in the larch forest. We modeled the A2G and G2G communication scenarios using a normal distribution. As illustrated in Fig.16 (a), the mean RMS-DS in the A2G channel is significantly larger, indicating more pronounced multipath effects compared to the G2G channel. Fig. 16 (b) presents the cumulative distribution function (CDF) curves for both A2G and G2G channels. The G2G curve lies above the A2G curve and exhibits a steeper slope, further confirming that the A2G channel experiences more significant multipath effects.

Table \uppercase\expandafter{\romannumeral5} outlines the mean RMS-DS values and the fitting parameters of the model in different forest environments. Typically, the mean value represents the average delay of the signal. A smaller mean indicates that the main energy of the signal arrives at the RX with a shorter delay, which may result in relatively weaker multipath effects. The variance reflects the dispersion of the signal delay. A larger variance suggests a wider distribution of signal delays, which usually indicates the presence of more multipath components. As shown in Table \uppercase\expandafter{\romannumeral5}, the mean and variance of RMS-DS in larch forests are smaller than those in birch forests, which implies that the multipath effects in larch forests are less pronounced than those in birch forests. Furthermore, Table \uppercase\expandafter{\romannumeral5} shows that the multipath effects in G2G communication are comparable to those in A2G scenarios with medium elevation angles. However, A2G scenarios, multipath effects are most pronounced at low elevation angles, primarily because the signal propagates over the longest distance through vegetation, particularly through dense tree canopies. At higher elevation angles, both the mean and the variance of the RMS-DS are minimized, primarily because the signal traverses a shorter distance through vegetation. Therefore, in forest environments, it is recommended to choose a 90 ° elevation angle for the A2G communication scenario, as this minimizes multipath interference. If signal coverage is a consideration, an elevation angle of no less than 60 ° should be preferred.

\subsection{Rician K Factor}

The Rician K-factor is defined as the ratio of the power of the direct path to the total energy of other multipath components. Fig. 17 illustrates the distribution of the Rician K factor for G2G and A2G channels within a larch forest at a 30 ° elevation angle. Both channels exhibit a normal distribution of the Rician K factor. As shown in Fig. 17 (b), the 90\% cumulative probability value for the A2G channel is 28.7, while for the G2G channel, it is 33.5. This indicates that the multipath effect is stronger in the A2G channel at a 30° elevation angle. The mean of the Rician K factor represents its typical value. A larger mean generally indicates more concentrated energy in the direct path and weaker multipath effects. The variance, on the other hand, reflects the fluctuation range of the Rician K factor, which is indicative of the complexity of the propagation environment. From Table \uppercase\expandafter{\romannumeral5}, it can be observed that the mean value of the Rician K factor in larch forests is larger, suggesting that the multipath effect in larch forests is relatively weak. In birch forests, however, the Rician K factor for the A2G channel at a 90° elevation angle is not the highest. This is due to the presence of several strong reflection paths near the main path in some data, which results in a negative Rician K factor and significantly reduces its mean value—a behavior distinct from that of RMS-DS.

Furthermore, Table \uppercase\expandafter{\romannumeral5} reveals that the multipath effect in high-elevation A2G scenarios is smaller than that in G2G scenarios, consistent with the trends observed for RMS-DS. Therefore, in forest environments, it is advisable to adjust the positions of the TX and RX to avoid high-energy multipath components. Alternatively, channel equalization algorithms can be implemented at the receiver to mitigate signal distortion caused by multipath propagation and enhance overall signal quality.



\section{Conclusion}
In this work, we conducted extensive G2G and A2G channel measurements using OFDM signals in the scenarios of deciduous larch and birch forests in the Arxan National Forest. Based on measurement data, we established propagation path loss models for G2G and A2G. These models are very consistent with the data and are suitable for mountainous forest environments with vegetation cover. Based on the directional antenna measurement results, the AS information was calculated. We analyzed the multipath effect in forest areas by extracting the CIR, calculated and statistically analyzed the RMS-DS and Rician K factor, and provided the modeled parameter results. The channel model proposed in this article can be applied to environmental sensing systems in forest areas, emergency rescue communication, etc. The parameters provided can guide channel simulation instruments. Due to the complex terrain of mountainous forest areas, UAVs can serve as mobile base stations to support additional traffic and improve positioning accuracy based on parameter information such as RMS-DS and Rician K factor provided in this article.In this study, the data collection and model validation were conducted during a specific period, which may not fully capture the effects of vegetation growth and seasonal variations. Seasonal changes, such as foliage growth in spring and summer or leaf fall in autumn, can significantly alter the scattering and absorption characteristics of vegetation. Some studies indicate that the channel variation in summer and winter can reach around 8 dB \cite{seasonal}. Our future work direction can measure forest channels in different seasons, and extend to the study of air-to-air channel propagation models, as well as the study of MIMO communication channel conditions in forest areas.

\end{document}